\documentclass[journal=jpcbfk,manuscript=article]{achemso}

\usepackage[version=3]{mhchem} 

\SectionNumbersOn

\usepackage{amsmath}
\usepackage{amssymb}
\usepackage{color}
\usepackage{CJK}
\usepackage{framed}
\usepackage{hyperref}
\usepackage{algorithm}
\usepackage{algpseudocode}
\algdef{SE}[DOWHILE]{Do}{doWhile}{\algorithmicdo}[1]{\algorithmicwhile\ #1}

\usepackage[scaled=0.85]{beramono}
\usepackage[T1]{fontenc}
\usepackage{changepage}


\newcommand*{\blauw}[1]{\textcolor{blue}{#1}}

\usepackage[draft]{todonotes}   
\usepackage{graphicx}

\author{Kirill Shmilovich}
\affiliation{%
  Pritzker School of Molecular Engineering, %
  University of Chicago, %
  Chicago, IL 60637, U.S.A.%
}

\author{Rachael A.\ Mansbach}
\affiliation{
	Theoretical Biology and Biophysics Group,  %
	Los Alamos National Laboratory,  %
	Los Alamos, NM 87545, U.S.A.
}

\author{Hythem Sidky}
\affiliation{%
  Pritzker School of Molecular Engineering, %
  University of Chicago, %
  Chicago, IL 60637, U.S.A.%
}

\author{Olivia E.\ Dunne}
\affiliation{%
  Pritzker School of Molecular Engineering, %
  University of Chicago, %
  Chicago, IL 60637, U.S.A.%
}

\author{Sayak Subhra Panda}
\affiliation{%
  Department of Chemistry,
  Johns Hopkins University,
  Baltimore, MD 21218, U.S.A.%
}
\alsoaffiliation{%
  Institute of NanoBioTechnology,
  Johns Hopkins University,
  Baltimore, MD 21218, U.S.A.%
}

\author{John D.\ Tovar}
\affiliation{%
  Department of Chemistry,
  Johns Hopkins University,
  Baltimore, MD 21218, U.S.A.%
}
\alsoaffiliation{%
  Institute of NanoBioTechnology,
  Johns Hopkins University,
  Baltimore, MD 21218, U.S.A.%
}
\alsoaffiliation{%
  Department of Materials Science and Engineering,
  Johns Hopkins University,
  Baltimore, MD 21218, U.S.A.%
}

\author{Andrew L.\ Ferguson}
\email{andrewferguson@uchicago.edu}
\affiliation{%
  Pritzker School of Molecular Engineering, %
  University of Chicago, %
  Chicago, IL 60637, U.S.A.%
}

\title[]{Discovery of Self-Assembling $\pi$-Conjugated Peptides by Active Learning-Directed Coarse-Grained Molecular Simulation}

\begin{document}

\begin{abstract}

\noindent Electronically-active organic molecules have demonstrated great promise as novel soft materials for energy harvesting and transport. Self-assembled nanoaggregates formed from $\pi$-conjugated oligopeptides composed of an aromatic core flanked by oligopeptide wings offer emergent optoelectronic properties within a water soluble and biocompatible substrate. Nanoaggregate properties can be controlled by tuning core chemistry and peptide composition, but the sequence-structure-function relations remain poorly characterized. In this work, we employ coarse-grained molecular dynamics simulations within an active learning protocol employing deep representational learning and Bayesian optimization to efficiently identify molecules capable of assembling pseudo-1D nanoaggregates with good stacking of the electronically-active $\pi$-cores. We consider the DXXX-OPV3-XXXD oligopeptide family, where D is an Asp residue and OPV3 is an oligophenylene vinylene oligomer (1,4-distyrylbenzene), to identify the top performing XXX tripeptides within all 20$^3$ = 8,000 possible sequences. By direct simulation of only 2.3\% of this space, we identify molecules predicted to exhibit superior assembly relative to those reported in prior work. Spectral clustering of the top candidates reveals new design rules governing assembly. This work establishes new understanding of DXXX-OPV3-XXXD assembly, identifies promising new candidates for experimental testing, and presents a computational design platform that can be generically extended to other peptide-based and peptide-like systems.
\end{abstract}

\clearpage
\newpage


\section{\label{sec:intro}Introduction}

Self-assembling $\pi$-conjugated peptides possessing a $\pi$-core flanked by peptide wings have emerged as a versatile building block for the bottom-up fabrication of bio-compatible nanoaggregates with engineered optoelectronic properties. Overlaps between $\pi$-orbitals in neighboring aromatic cores within supramolecular assemblies lead to the emergence of optical and electronic properties including fluorescence, electron/hole transport,
and exciton splitting, and the flanking oligopeptide wings provide the capacity to operate in and interact with biological environments \cite{Pinotsi2016ProtonStructures,Guo2013DesigningElectronics, kim2012model, mitschke2000electroluminescence, roncali1992conjugated, fichou1999structure, Bian2012RecentCells,  guo2013designing, newman2004introduction, marder2008photoresponsive, beaujuge2010color, marty2013hierarchically}. These peptidic materials have proven readily synthesizable and responsive to external control mediated by pH, flow, light, salt concentration, and temperature \cite{Lowik2010StimulusMaterials,Ulijn2008DesigningNanomaterials,Marciel2013Fluidic-DirectedCores,Webber2016SupramolecularBiomaterials,Mansbach2017ControlFlow,Schenning2005SupramolecularSystems,Mba2011SynthesisWater,Gallaher2012ControlledPerylene-bisimides}, and have found a host of potential applications in the context of photovoltaic power generation, energy harvesting, and as organic transistors \cite{Facchetti2011-Conjugatedsup/sup,Bian2012RecentCells,Wall2014SupramolecularSequence,newman2004introduction,Beaujuge2010ColorDevices,Ardona2015EnergyEnvironments,Thurston2016ThermodynamicsOligopeptides,Sanders2017Solid-PhasePeptides,Valverde2018EvidenceSpectroscopy}. The structural and functional properties of the self-assembled nanoaggregates are governed by the molecular chemistry of the $\pi$-core and the amino acid sequence of the peptide wings. 

\begin{figure*}[ht!]
 \centering
 \includegraphics[width=\textwidth]{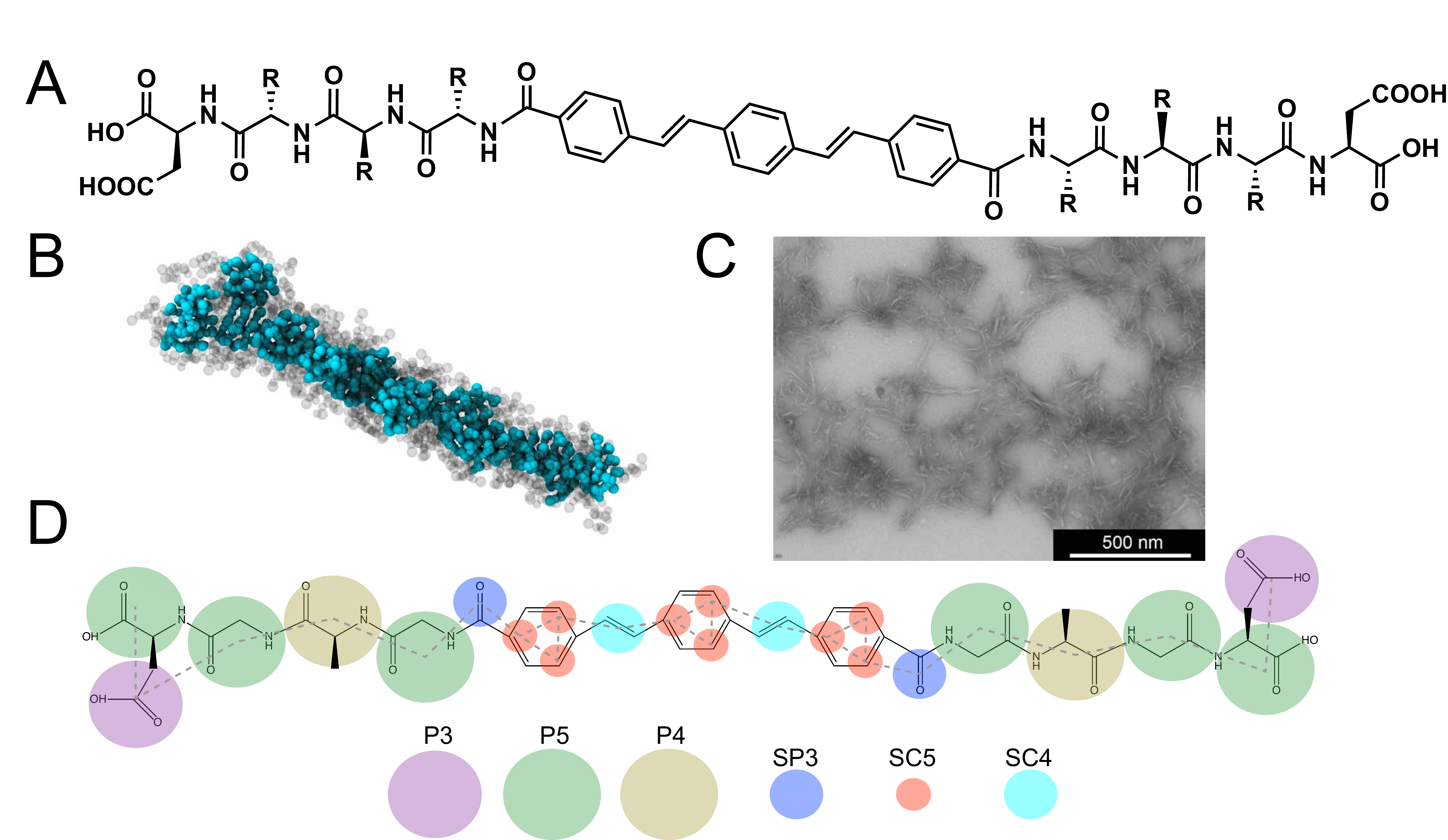}
 \caption{The DXXX-OPV3-XXXD system. (a) Chemical structure of the prototypical DXXX-OPV3-XXXD peptide monomer. The oligophenylenevinylene $\pi$ core (OPV3) is flanked by oligopeptide wings (DXXX) that are mirror symmetric such that the identity of the amino acids is inverted and the molecule possesses two C-termini. The X residues are selected from the 20 natural amino acids such that the family comprises 20$^3$ = 8,000 distinct molecules. (b) Molecular simulation snapshot of a self-assembled pseudo-1D nanoaggregate spontaneously formed by the spontaneous association of DVAA-OPV3-VAAD peptide monomers into a linear stack. Good stacking between the $\pi$-cores (colored blue) favors $\pi$ orbital overlap, electronic delocalization along the backbone of the nanoaggragate, and the emergence of optical and electronic functionality. (c) Experimental transmission electron microscopy (TEM) image of self-assembled fibrils formed by DFFG-OPV3-GFFD peptides in an acidic environment. Reprinted with permission from Ref.~\cite{Wall2014SupramolecularSequence}\ . Copyright (2014) American Chemical Society. (d) Illustration of the mapping from the DGAG-OPV3-GAGD all-atom structure to the coarse-grained representation at which the simulations in this work are conducted. The coarse-grained beads corresponding to groupings of neighboring atoms are labeled according to the Martini model employed in this work \cite{SiewertJ.Marrink2007TheSimulations,monticelli2008martini,deJong2013ImprovedField}.}
 \label{fgr:peptide}
\end{figure*}

The Asp-X-X-X-(oligophenylenevinylene)$_3$-X-X-X-Asp (DXXX-OPV3-XXXD) family represents one class of synthetic $\pi$-conjugated peptides possessing an oligophenylenevinylene $\pi$ core, terminal Asp residues, and amino acid side chains, where X represents one of the 20 natural amino acids (Fig.~\ref{fgr:peptide}a). To assure the molecules are head-to-tail invariant, the oligopeptide wings are constrained to be mirror-symmetric both in the identity of the amino acids and the N-to-C directionality, such that each molecule possesses two C-termini. The terminal residues are constrained to be Asp to endow each terminus of the molecule with two carboxyl groups and provide a pH trigger for assembly: at pH>5 the four carboxyls are deprotonated endowing the molecule with a (-4)$e$ formal charge and disfavoring large scale assembly, but at pH<1 the residues protonate, the molecule becomes neutral, and large-scale aggregation proceeds~\cite{Valverde2018EvidenceSpectroscopy}. The DXXX-OPV3-XXXD family has attracted considerable experimental and computational attention in recent years due to their demonstrated capability to assemble into pseudo-1D optically and electronically active nanoaggregates whose structure and properties can be tuned through selection of the X residues \cite{thurston2019revealing,Mansbach2017,Mansbach2017ControlFlow,Thurston2016ThermodynamicsOligopeptides,Wall2014SupramolecularSequence,Wall2012SynthesisMaterials}. Assembly in aqueous solvent under acidic conditions is driven by hydrophobic, $\pi$-$\pi$ stacking, and hydrogen bonding interactions \cite{Thurston2016ThermodynamicsOligopeptides,thurston2019revealing,Thurston2018MachineOligopeptides,Valverde2018EvidenceSpectroscopy}. The assembly of elongated peptides into linear aggregates with in-register stacking and alignment of the $\pi$-cores favors $\pi$ orbital overlap, electronic delocalization along the backbone of the nanoaggragate, and the emergence of optical and electronic functionality such as well-defined absorption and emission spectra, HOMO/LUMO gaps, electron/hole conductivity, and exciton splitting capabilities (Fig.~\ref{fgr:peptide}b,c) \cite{thurston2019revealing, Wall2014SupramolecularSequence, wall2011aligned, panda2018solid, kumar2011hierarchical, panda2019controlling}.

The complete DXXX-OPV3-XXXD family comprises 20$^3$ = 8,000 members corresponding to all possible permutations of the 20 natural amino acids within the unspecified XXX triplet. This vast size of this chemical space is both a blessing -- the large palette of molecular chemistries provides enormous versatility in materials properties and the opportunity to tailor structure and function -- and a curse -- it is a challenge to identify promising candidates within this enormous space. Identifying the candidates capable of self-assembling into well-ordered optoelectronic nanoaggregates and divining the design precepts dictating the mechanism is a key goal in realizing these peptides as novel biocompatible optoelectronic materials.

Edisonian traversal of the large chemical space of DXXX-OPV3-XXXD molecules by trial-and-improvement experimentation is essentially intractable due to the high time and labor costs associated with peptide synthesis and testing. To date, no more than 13 members of the family have been experimentally synthesized and tested \cite{Wall2014SupramolecularSequence}. Molecular simulation offers an alternative means to perform high-throughput virtual screening of chemical space to identify the most promising candidates for experimental testing. Since assembly proceeds on length scales of tens of nanometers and microsecond time scales, this has motivated the development of coarse-grained models explicitly parameterized against all-atom molecular simulations \cite{Mansbach2017ControlFlow,Mansbach2017,Mansbach2018PatchyPeptides} (Fig.~\ref{fgr:peptide}d). These models integrate out the electronic and atomistic degrees of freedom by lumping together small numbers of atoms into beads in order to furnish a molecular model that offers a judicious compromise between chemical realism and the computational efficiency required to directly simulate peptide assembly \cite{Mansbach2017}. Exhaustive simulation of all 8,000 candidates within the DXXX-OPV3-XXXD family remains, however, computationally expensive. As we shall demonstrate, however, doing so is unnecessary to parameterize a reliable surrogate model of peptide function and identify and validate the most promising candidates within the family.

Chemical intuition is extremely valuable in guiding the computational search through chemical space, but it can perform poorly in the limits of data paucity, where there are too few examples to infer patterns, and data abundance, where there are too many examples to parse effectively. Further, inherent preconceptions and biases may push the search away from potentially profitable regions of chemical space and overlook patterns in the high-dimensional data that may reveal important determinants of molecular performance. Active learning (a.k.a.\ sequential learning, optimal experimental design), and more specifically, Bayesian optimization, presents a systematic approach to guide traversal of chemical space by using information on all measurements conducted to date to inform the ``next-best'' measurement to conduct \cite{kim2019active,ling2017high,barrett2019iterative,Gomez-Bombarelli2018AutomaticMolecules,Bickerton2012QuantifyingDrugs}. In this manner, active learning predicts an optimal sequence in which to consider the molecular candidates in order to identify the optimal ones with minimal data collection effort. For this reason, active learning and allied approaches have been rapidly gaining traction in the materials discovery, engineering, and design communities, with these approaches being deployed, for example, in the experimental discovery of novel shape memory alloys \cite{xue2016accelerated}, piezoelectrics \cite{yuan2018accelerated}, high glass transition polymers \cite{kim2019active}, the computational discovery of drugs \cite{Gomez-Bombarelli2018AutomaticMolecules}, and magnetocaloric, superconducting, and thermoelectric materials \cite{ling2017high}.

Our primary goal is to efficiently identify members of the DXXX-OPV3-XXXD family that exhibit self-assembly into desired pseudo-1D nanoaggregates with good overlap between the $\pi$-conjugated cores and are thus most promising in displaying emergent optical and electronic functionality. We adopt a coarse-grained bead-level molecular simulation model as the engine for our high-throughput virtual screen and couple this with a deep learning-enabled active learning protocol to guide optimal traversal of chemical space. We identify and computationally validate the top performing constituents of the 8,000-member DXXX-OPV3-XXXD family after simulating only 2.3\% of all possible molecules. This represents a massive saving over exhaustive sampling enabled by active learning. The absence of any introduced human bias within the active learning protocol also proved to be valuable in identifying high-performing candidates incorporating methionine residues that were not previously considered. A \textit{post hoc} analysis of the observed assembly pathways provides supporting mechanistic understanding of the self-assembly behavior and exposes practical precepts for molecular design. The rank ordered list of DXXX-OPV3-XXXD molecules produced by our computational analysis provides a useful filtration of the design space with the top-performing candidates offering a massively reduced candidate space for experimental synthesis and testing.

\section{\label{sec:methods}Methods}

\subsection{Molecular dynamics simulation} \label{sec:MD}

The DXXX-OPV3-XXXD peptides were modeled using a previously-developed coarse-grained potential based on the Martini potential \cite{Mansbach2017,Mansbach2017ControlFlow}. Martini is a popular coarse-grained potential that lumps approximately four heavy atoms into each coarse-grained bead, has demonstrated great successes in modeling peptides, proteins, lipids, and carbohydrates \cite{monticelli2008martini, deJong2013ImprovedField, gautieri2010coarse, pannuzzo2014simulation, lopez2013computational, guo2012probing, seo2012improving}, and offers a good compromise between chemical specificity and the computational efficiency necessary to probe the formation of large peptide aggregates. The potential was initially developed for DFAG-OPV3-GAFD by refitting the native Martini parameters for bonded interactions against all-atom simulation data \cite{Mansbach2017}. This bottom-up reparameterization of the bonded interactions greatly improved agreement between the coarse-grained and all-atom distribution functions, potentials of mean force (PMF) for monomer stretching and dimerization, and time-averaged contact maps \cite{Mansbach2017}. We generalize this model to the complete DXXX-OPV3-XXXD family by maintaining the same parameterization of the bonds, angles, and backbone dihedrals within the OPV3 core and employing default Martini parameters for the amino acid side chains and all non-bonded interactions \cite{SiewertJ.Marrink2007TheSimulations,deJong2013ImprovedField}. An illustration of the all-atom to coarse-grained bead-level mapping for DGAG-OPV3-GAGD is provided in Fig.~\ref{fgr:peptide}d. Calculation and comparison of the translational diffusion constants for the all-atom and coarse-grained models of DFAG-OPV3-GAFD showed these to be in agreement within error bars, indicating no significant discrepancy in the (translational) dynamical time scales between the two models and that no time scale corrections to the coarse grained calculations are required.

Coarse-grained molecular dynamics simulations of peptide assembly were conducted using the Gromacs 2018.6 simulation suite \cite{Abraham2015GROMACS:Supercomputers}. Initial system configurations for each DXXX-OPV3-XXXD considered were generated by randomly inserting 96 peptides into a 16.2$\times$16.2$\times$16.2 nm$^3$ cubic simulation box with 3D periodic boundary conditions, corresponding to a concentration of approximately 35 mM. The amino acid residues are prepared in protonation states corresponding to pH 1 to mimic pH-triggered experimental assembly under acidic conditions. The coarse-grained peptides were then solvated in water to a density of 1.0 g/cm$^3$ of water using the Martini non-polarizable water model \cite{SiewertJ.Marrink2007TheSimulations}. Steepest descent energy minimization was performed to eliminate high energy overlaps by removing forces greater than 1,000 kJ/mol.nm. Initial particle velocities were assigned from a Maxwell-Boltzmann distribution at 298 K. All simulations were conducted in the NPT ensemble at 298 K and 1 bar using a velocity-rescaling thermostat \cite{Bussi2007CanonicalRescaling} and Parrinello-Rahman barostat \cite{Parrinello1981PolymorphicMethod}. Equations of motion were numerically integrated using the leap-frog algorithm with a 5 fs time step \cite{Hockney1988ComputerParticles} and bond lengths fixed using the LINCS algorithm \cite{Hess1997LINCS:Simulations}. Lennard Jones interactions were smoothly shifted to zero at 1.1 nm and reaction-field electrostatics were employed using a relative electrostatic screening constant of 15 appropriate for the non-polarizable water model \cite{monticelli2008martini}. An initial 100 ps equilibration run was conducted, after which time the temperature, pressure, density, and energy all stabilized. This was followed by a 3 $\mu$s production run, after which time the structural evolution of the system as measured by graphical analysis of the self-assembled aggregate (see Section \ref{sec:ObjFunc}) was stationary in time. Simulation snapshots were harvested for analysis every 50 ps over the course of the production run. Calculations were predominantly conducted on single NVIDIA GeForce RTX 2080 Ti cards and achieved execution speeds of $\sim$1.45 $\mu$s/day.

\subsection{Active learning peptide discovery} \label{sec:activeLearning}

\begin{figure*}[ht!]
 \centering
 \includegraphics[width=0.95\textwidth]{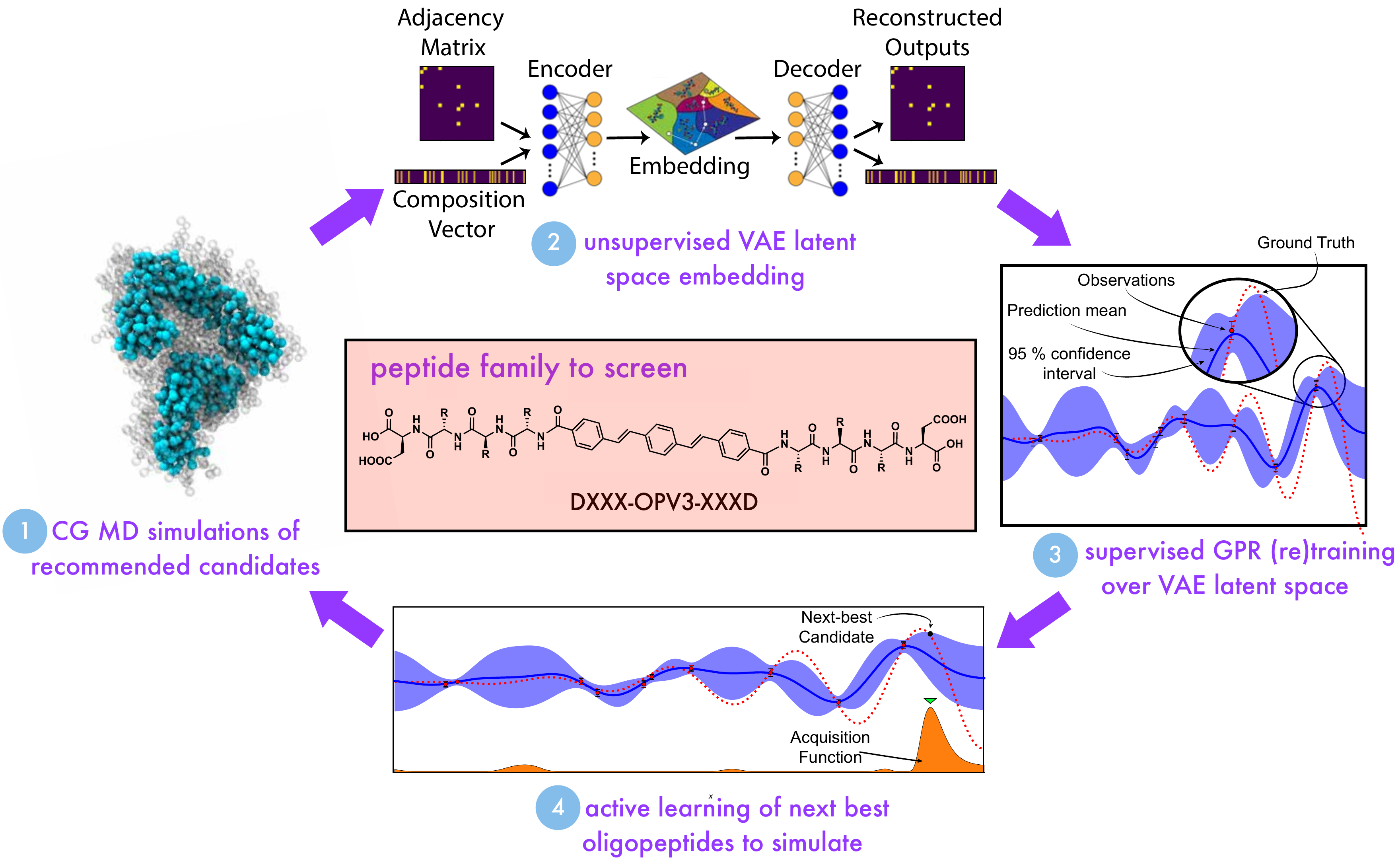}
 \caption{Active learning cycle for the data-driven discovery of optimally self-assembling DXXX-OPV3-XXXD peptides. The cycle contains four components. (1) Coarse-grained molecular simulations are performed on selected DXXX-OPV3-XXXD candidates and the quality of the self-assembled aggregates formed by molecule $i$ measured according to a scalar fitness $f_i$. (2) The DXXX-OPV3-XXXD family is projected from the high-dimensional chemical space of molecular structures into a low-dimensional latent space embedding $E : \text{DXXX-OPV3-XXXD}_i \rightarrow \mathbf{z}_i \in \mathbb{R}^d$ using a variational autoencoder (VAE). The dimensionality of the latent space is optimized during each cycle. (3) A Gaussian process regression (GPR) model is constructed over the VAE latent space linking the latent space coordinates of each DXXX-OPV3-XXXD family member to the scalar fitness function measuring the quality of their self-assembled aggregates $f : \hat{f}_i = f(\mathbf{z}_i) = (f \circ E)(\text{DXXX-OPV3-XXXD}_i)$. The GPR mapping $f$ is retrained each cycle over all DXXX-OPV3-XXXD candidates that have been simulated to date and for which measures of the fitness function $f_i, \; i \in sampled$ is available, and is then used to predict the fitness of unsimulated candidates $\hat{f}_j, \; j \in unsampled$. (4) The predicted means and uncertainties for $\hat{f}_j, \; j \in unsampled$ furnished by the GPR surrogate model are combined within an active learning acquisition function to identify the ``next-best'' candidates for which to perform coarse-grained molecular simulations to drive sampling towards the most promising candidates. The loop is cycled until the GPR surrogate model no longer changes with additional data collection and can then be used to reliably identify the top candidates for computational validation.}
 \label{fgr:Pipeline}
\end{figure*}

An active learning protocol is employed to direct a principled traversal of the DXXX-OPV3-XXXD candidate space and minimize the number of coarse-grained simulations required to discover the highest-performing candidates \cite{kim2019active,ling2017high,barrett2019iterative}. The fundamental challenge is that evaluating the quality of each peptide by direct simulation is expensive, so we wish to identify the best peptide candidates in the fewest number of simulations. The procedure we employ is in large part inspired by and adapted from a pioneering deep representational active learning approach for molecular drug discovery developed by Gomez-Bombarelli et al. \cite{Gomez-Bombarelli2018AutomaticMolecules}. Our approach comprises four main steps and is illustrated schematically as an iterative active learning cycle in Fig.~\ref{fgr:Pipeline}. The coarse-grained molecular simulation engine representing our measurement function within the protocol is described in Section \ref{sec:MD}, and we define our fitness function in Section \ref{sec:ObjFunc}. Appreciating that some of the more technical machine learning concepts may be foreign to some readers in the molecular modeling community, we expose these steps in the protocol in some detail along with their specific adaptations to our molecular system, but those readers familiar with variational autoencoders, Gaussian process regression, and Bayesian optimization may feel free to skim over Sections \ref{sec:ChemEmbed}-\ref{sec:stop}. All codes are developed in Python 3 making use of the Scikit-learn~\cite{scikit-learn}, NumPy~\cite{numpy}, Keras~\cite{chollet2015keras}, and ORCA \cite{Hocevar2014ACounting} libraries. Jupyter notebooks implementing our methods are hosted on GitHub (\url{https://github.com/KirillShmilovich/ActiveLearningCG}).

\subsubsection{Step 1: Definition of fitness function for self-assembled aggregates} \label{sec:ObjFunc}

To perform active learning discovery in our predefined chemical space we define a scalar-valued fitness function $f : f_i = f(\text{DXXX-OPV3-XXXD}_i)$ that assigns a quality to each peptide in terms of its capacity to self-assemble into pseudo-1D nanoaggregates. Linear aggregates with good overlap between the $\pi$-conjugated cores are most promising in displaying emergent optical and electronic functionality and therefore anticipated to possess the most desirable materials properties. We have previously employed DFT calculations to make direct predictions of optoelectronic properties, but the high computational cost of these calculations limit them to aggregates of small numbers of peptides (dimers and trimers) and require omission of the flanking amino acid residues and solvent \cite{thurston2019revealing}. As such, these calculations are poorly suited to high-throughput virtual screening for large-scale aggregation behavior. Consequently, we define and optimize a structural measure of assembly quality in our coarse-grained molecular simulations as a proxy for optical and electronic functionality. This simplification massively expedites sampling in the full chemical space and provides a means to coarsely screen chemical space and focus a subsequent experimental search on the most promising candidates. Alternatively, this computational screen can be viewed as a preliminary filtration within the coarsest level of a nested hierarchy of increasingly expensive all-atom and/or electronic structure calculations.

In order to specify $f$ we define a geometric criterion by which a pair of peptides are considered to form part of the same pseudo-1D nanoaggregate. To do so, we adopt a distance metric that we have previously employed to define clustering in DFAG-OPV3-GAFD assembly \cite{Mansbach2017,Mansbach2017ControlFlow} and asphaltene aggregation \cite{Wang2016MesoscaleAggregation}. This so-called ``optical distance'' metric is defined as the minimum center of mass distance between aromatic cores in molecule $a$ and $b$, 
\begin{equation}
    d^\text{optical}_{ab} = \min _{i \in core(a), j \in core(b)} r_{ij}, \label{eqn:dopt}
\end{equation}
where $r_{ij}$ is the intermolecular center-of-mass distance between the aromatic rings $i$ and $j$ within the OPV3 cores, and the minimization proceeds over the three aromatic rings $i \in core(a)$ in molecule $a$, and the three aromatic rings $j \in core(b)$ in molecule $b$. Pairs of molecules $a$ and $b$ which satisfy $d^\text{optical}_{ab} < r_\text{cut} = 0.7$ nm are considered to reside within the same cluster \cite{Mansbach2017,Mansbach2017ControlFlow}. The cutoff $r_\text{cut} = 0.7$ was tuned to the mean of the distribution of $d^\text{optical}_{ab}$ collected over DFAG-OPV3-GAFD peptide dimers with good in-register stacking of the OPV3 cores \cite{Mansbach2017}. In contrast with other choices of peptide clustering metrics based, for example, on the overall center-of-mass or proximity of the peptide wings, the optical metric assures close intermolecular proximity of at least one pair of OPV3 aromatic rings in a pair of associated peptides. This close association promotes $\pi$ electron overlap, electron delocalization, and the emergence of optoelectronic function, and it is for this reason that this metric is termed the optical distance metric \cite{Mansbach2017,Mansbach2018PatchyPeptides,Mansbach2017ControlFlow}.

\begin{figure}[ht!]
\centering
  \includegraphics[width=\linewidth]{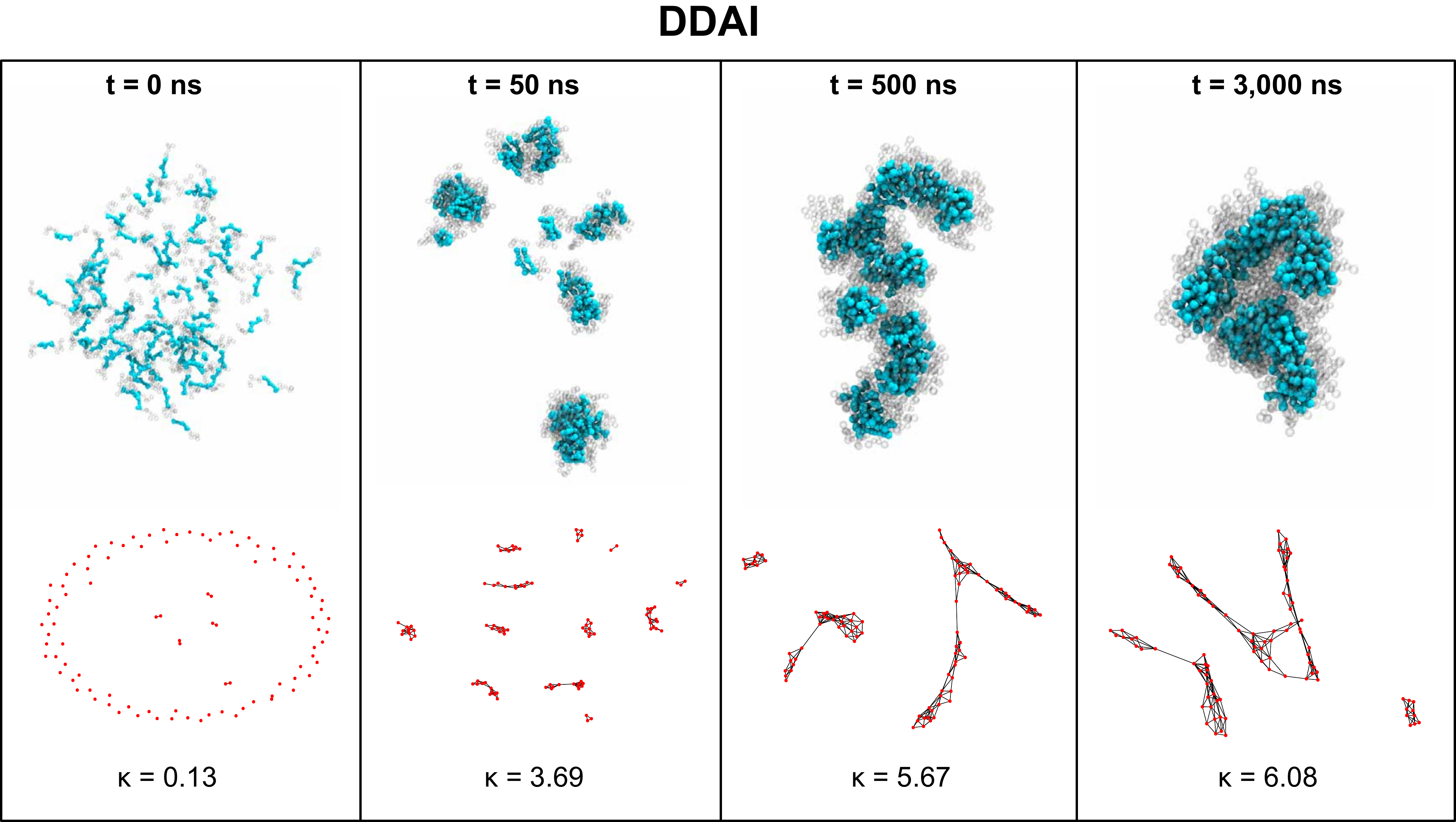}
  \caption{Dynamical evolution of the self-assembled structures of DDAI-OPV3-IADD over the course of a 3 $\mu$s coarse-grained molecular simulation. Snapshots of the molecular simulation show molecules in which the OPV3 $\pi$ cores are colored blue, the peptide wings faded grey, and water is removed for clarity. The interaction graph corresponding to each snapshot is shown directly below each image. The vertices $V$ corresponding to each peptide are colored red, and edges $E$ between peptide pairs defined by $d^\text{optical}_{ab} < r_\text{cut} = 0.7$ (Eqn.~\ref{eqn:dopt}) and colored grey. The average degree $\kappa = \frac{2 |E|}{|V|}$ is reported at the bottom of each panel. At $t$ = 0 ns, the 96 randomly placed peptides form essentially a monomeric dispersion, with the exception of five dimer pairs, and the system possesses a correspondingly low $\kappa$ = 0.13. As the simulation progresses, the peptides self assemble under the influence of hydrogen bonding, $\pi$-$\pi$ stacking, and hydrophobic interactions into small ($t$ = 50 ns, $\kappa$ = 3.69) and then larger ($t$ = 500 ns, $\kappa$ = 5.67; $t$ = 3,000 ns, $\kappa$ = 6.08) aggregates with a commensurate increase in the mean degree $\kappa$. In this figure, and throughout the paper, molecular renderings are generated using VMD \cite{Humphrey1996VMD:Dynamics.}, and interaction graphs produced using NetworkX \cite{hagberg2008exploring}.}
  \label{fgr:Graphs}
\end{figure}

Given this definition, a natural choice for the fitness $f_i$ of molecule $i$ is the number of such optical contacts in a self-assembled aggregate, since maximizing this value will promote electronic delocalization and the emergence of optoelectronic functionality. We evaluate the fitness function by representing the molecular system as a dynamically-evolving interaction graph in which the peptides compose the vertices $V = \{v_1, v_2 ,\ldots, v_N\}$ and the edges $E = \{e_{1,2}, e_{1,3}, ... , e_{95,96}\}$ are assigned between pairs of vertices $v_a$ and $v_b$ if $d^\text{optical}_{ab} < r_\text{cut} = 0.7$. An illustration of the evolution of the interaction graph over the course of a 3 $\mu$s simulation of DDAI-OPV3-IADD assembly is presented in Fig.~\ref{fgr:Graphs}. The number of vertices $|V|$ = 96 is fixed by the number of peptides in the system. Maximization of the number of edges $|E(t)|$ at time $t$ is therefore equivalent to maximizing the mean degree of each vertex in the graph $\kappa(t) = \frac{2 |E(t)|}{|V|}$. As such, we adopt as our fitness function,
\begin{align}
    f_i &= \overline{\kappa(t; \text{DXXX-OPV3-XXXD}_i)} = \frac{2 \overline{|E(t; \text{DXXX-OPV3-XXXD}_i)|}}{|V|}, \label{eqn:kappa}
\end{align}
where the time average denoted by the overbar is performed over the terminal 50 ns of the 3 $\mu$s production run. Standard errors in the mean are estimated by block averaging the terminal 50 ns in five contiguous 10 ns blocks.

A potential criticism of the fitness function is that $\kappa$ achieves a maximum for all-to-all connectivity of the graph, and its maximization would therefore appear to not necessarily favor pseudo-1D linear stacks. Mathematically this is true, but there are strong physical limitations on the maximum attainable value of $\kappa$ since the excluded volume of the $\pi$ cores allow then to form optical associations with a limited number of partners. The largest value observed in all of our calculations is $\kappa$ = 6.07 (cf.~Table \ref{tbl:BestDXXX}), and visual inspection of the terminal aggregates confirms that $\kappa$ is positively correlated with the formation of elongated pseudo-1D nanoaggregates similar to those illustrated in the $t$ = 3,000 ns panel of Fig.~\ref{fgr:Graphs}.

\subsubsection{Step 2: Learning latent space embeddings using variational autoencoders} \label{sec:ChemEmbed}

In Step 3 (Section \ref{sec:GPR}) we describe our training of a Gaussian process regression (GPR) surrogate model to predict the fitness of candidate molecules that have not been simulated based on those that have. The predictions of this model are then used to perform active learning. We experimented with constructing the GPR directly over the chemical space of DXXX-OPV3-XXXD molecules by measuring pairwise distances between the XXX tripeptides using BLOSUM substitution matrices \cite{henikoff1992amino}, but following Gomez-Bombarelli et al.~\cite{Gomez-Bombarelli2018AutomaticMolecules}, we found this approach to yield inferior surrogate models to those constructed over data-driven low-dimensional embeddings of the molecules generated using a variational autoencoder (VAE) \cite{Kingma2013Auto-EncodingBayes}. The low-dimensional VAE latent spaces also conveys advantages in that low-dimensional GPRs tend to be more robust, chemically similar molecules tend to be embedded proximately in the latent space providing interpretability of the chemical space through dimensionality reduction, and the continuous and differentiable nature of the latent space makes it well-suited to global optimization~\cite{Gomez-Bombarelli2018AutomaticMolecules,Calandra2014ManifoldRegression}.

\begin{figure}[ht!]
\centering
  \includegraphics[width=0.8\linewidth]{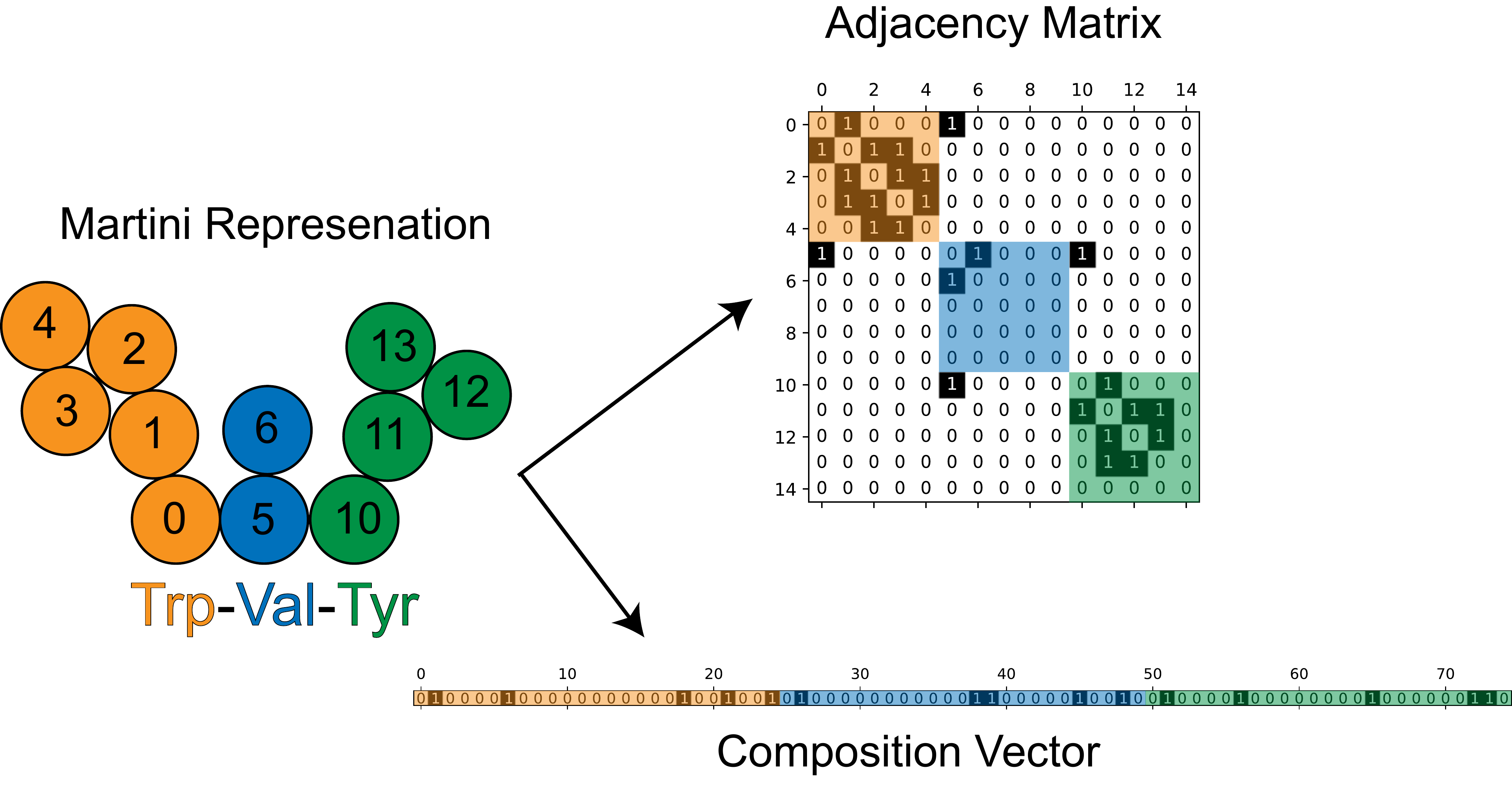}
  \caption{Schematic of the representation of each XXX tripeptide to the VAE. The Martini representation of each tripeptide $i$, in this example Trp-Val-Tyr (WVY), is converted into an adjacency matrix $\mathbf{A}_i \in \mathbb{R}^{15\times15}$ specifying the connectivity of beads within the tripeptide and a one-hot composition vector $\mathbf{T}_i \in \mathbb{R}^{75}$ specifying the identity of the beads. The tuple $(\mathbf{A}_i,\mathbf{T}_i)$ defines the input provided to the VAE. A pre-defined sequential numbering is employed for the beads in each amino acid. The adjacency matrix is padded with rows and columns of zeros to represent tripeptides containing fewer than the maximum number of 15 beads. The colored blocks in the adjacency matrix and composition vector correspond to the colors of the amino acids in the Martini molecule.}
  \label{fgr:AdjMat}
\end{figure}

We represent the DXXX-OPV3-XXXD molecules to the VAE only through the identity of the XXX tripeptide, since this is the only differentiating feature between molecules. We base this representation on the coarse-grained Martini model used to perform our molecular simulations. This representation comprises two components for each molecule $i$: (i) an adjacency matrix $\mathbf{A}_i$, which captures the connectivity of beads within the tripeptide, and (ii) a one-hot encoded composition vector of bead-types $\mathbf{T}_i$ specifying the identity of the Martini beads (Fig.~\ref{fgr:AdjMat}). Since peptide sequences may contain varying numbers of coarse-grained beads, we standardize the size of the adjacency matrix $\mathbf{A}_i \in \mathbb{R}^{15\times15}$ to be sufficiently large enough to accommodate the largest tripeptide (Trp-Trp-Trp) and pad the array with zeroes for smaller molecules. A one-hot composition vector of length $\mathbf{T}_i \in \mathbb{R}^{75}$ is sufficient to accommodate all tripeptide compositions considered. For each molecule $i$, the tuple $(\mathbf{A}_i,\mathbf{T}_i)$ defines the input provided to the VAE.

\begin{figure*}[ht!]
 \centering
 \includegraphics[width=\textwidth]{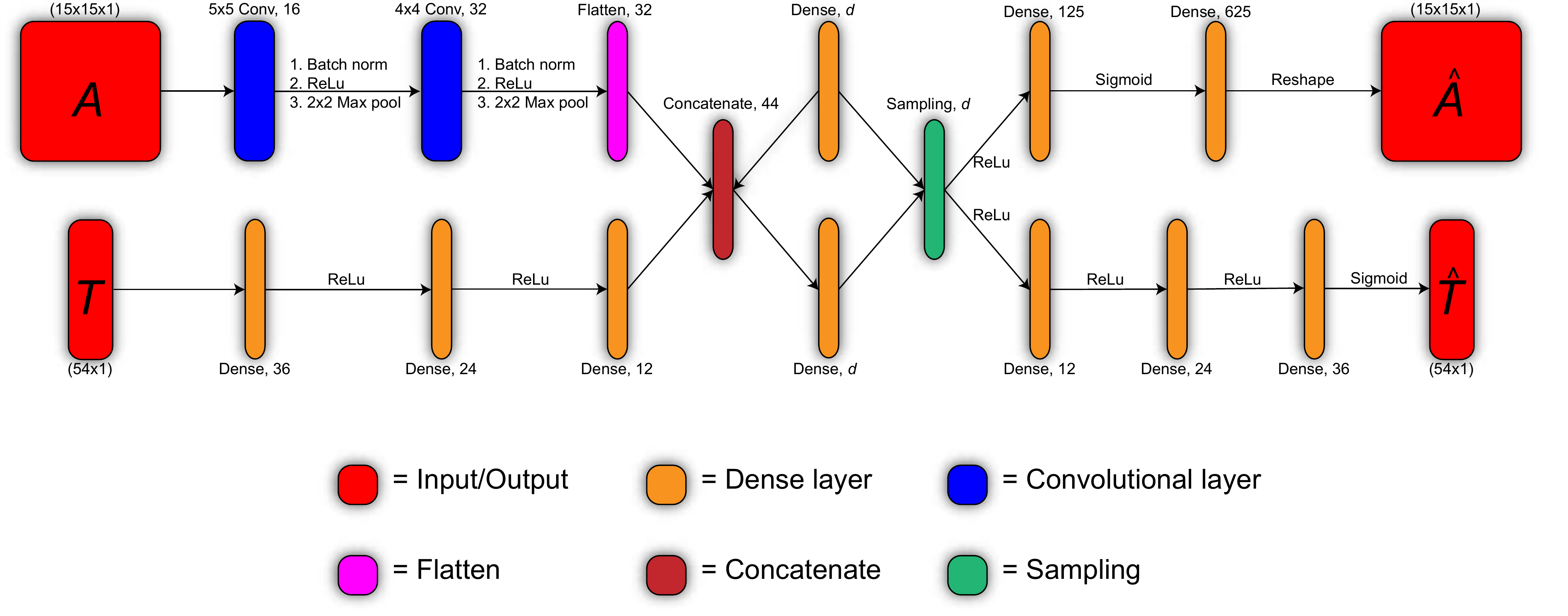}
 \caption{Architecture of the variational autoencoder (VAE) used to generate the DXXX-OPV3-XXXD latent space embedding. The VAE accepts as inputs adjacency matrix and composition vector tuples $(\mathbf{A}_i,\mathbf{T}_i)$ and employs two parallel encoders to perform feature extraction and learn the mean $\boldsymbol{\mu}_i$ and standard deviation $\boldsymbol{\sigma}_i$ of a Gaussian distributed latent space embedding $\mathbf{z}_i \sim \mathcal{N}(\boldsymbol{\mu}_i ,\boldsymbol{\sigma}_i) \in \mathbb{R}^{d}$. The decoder generates approximate reconstructions $(\hat{\mathbf{A}}_i, \hat{\mathbf{T}}_i)$ of the inputs from the latent space representation. The network is trained by minimizing a loss function balancing reconstruction accuracy and a regularization term constraining the latent space to follow a multidimensional Gaussian distribution (Eqn.~\ref{eqn:loss}). The dimensionality $d$ of the latent space is treated as a hyperparameter that is optimized during each cycle of active learning.}
 \label{fgr:VAE}
\end{figure*}

The architecture of the VAE is illustrated in Fig.~\ref{fgr:VAE}. Given the two-part input $(\mathbf{A}_i,\mathbf{T}_i)$ for molecule $i$, the encoder block processes this through two parallel networks to perform feature extraction from each input. The $\mathbf{A}_i$ resemble a small image which motivate using a short series of convolutional layers to treat these inputs, whereas the binary $\mathbf{T}_i$ vectors are passed through a series of fully-connected dense layers. The features extracted by the encoder through the two parallel encoder networks are subsequently concatenated and used to generate the mean $\boldsymbol{\mu}_i$ and standard deviation $\boldsymbol{\sigma}_i$ of a Gaussian distributed latent space embedding $\mathbf{z}_i \sim \mathcal{N}(\boldsymbol{\mu}_i , \boldsymbol{\sigma}_i) \in \mathbb{R}^{d}$. The dimensionality of the latent space is treated as a hyperparameter that is optimized during each cycle of active learning and is found to lie in the range $d \in [4,10]$. The decoder then attempts to reconstruct $(\mathbf{A}_i,\mathbf{T}_i)$ from the latent encoding $\mathbf{z}_i$ again using two parallel networks. The part of the decoder predicting the reconstruction $\hat{\mathbf{T}}_i$ is identical to the architecture of the encoder, whereas the part predicting the reconstruction $\hat{\mathbf{A}}_i$ is simply another series of fully-connected layers that is reshaped to match the size of the input. The overall action of the VAE is the functional composition of the encoder $E : \mathbf{z}_i = E(\mathbf{A}_i,\mathbf{T}_i)$ and decoder $D : (\hat{\mathbf{A}}_i,\hat{\mathbf{T}}_i) = D(\mathbf{z}_i)$ blocks such that the total effect of the network is $(\hat{\mathbf{A}}_i,\hat{\mathbf{T}}_i) = (D \circ E)(\mathbf{A}_i,\mathbf{T}_i)$.

The VAE is trained by minimizing the VAE loss $\mathcal{L}_{VAE}$ composed of a reconstruction term $\mathcal{L}_{Rec}$ and a Kullback-Leibler (KL) divergence term $\mathcal{L}_{KL}$ \cite{Kingma2013Auto-EncodingBayes,doersch2016tutorial},
\begin{align}
    \mathcal{L}_{VAE} &= \mathcal{L}_{Rec} + \mathcal{L}_{KL}, \label{eqn:loss} \\ 
    \mathcal{L}_{Rec} &= \mathbb{E}_{i}[\text{BCE}(\hat{\mathbf{A}}_i, \mathbf{A}_i) + \text{BCE}(\hat{\mathbf{T}}_i, \mathbf{T}_i)]) \nonumber \\
    &\approx \sum_{i \in \text{mini-batch}}[\sum_{j=1}^{15}(A_{i,j}\log(\hat{A}_{i,j}) + (1-A_{i,j})\log(1-\hat{A}_{i,j})) \nonumber \\
    &\qquad+ \sum_{j=1}^{75}(T_{i,j}\log(\hat{T}_{i,j}) + (1-T_{i,j})\log(1-\hat{T}_{i,j}))], \\
    \mathcal{L}_{KL}  &= -D_{KL}(\mathbf{z} = E(\mathbf{A}, \mathbf{T})~||~\mathcal{N}(0, \mathbf{I})) \nonumber \\
    &\approx \sum_{i \in \text{mini-batch}}[-\frac{1}{2}\sum_{j=1}^d(1 + \log(\sigma_{i,j}^2) - \sigma_{i,j}^2 - \mu_{i,j}^2)],
\end{align}
where $\text{BCE}(\mathbf{x}, \mathbf{y})$ is the binary cross entropy between the reconstructions $\mathbf{x}$ and ground truth $\mathbf{y}$, $D_{KL}(Q~||~P)$ is the Kullback-Leibler divergence from $P$ to $Q$, and $\mathbf{I}$ is the $n$-by-$n$ identity matrix. The reconstruction term $\mathcal{L}_{Rec}$ encourages the VAE to reconstruct the inputs through the low-dimensional latent space information bottleneck. In contrast to a vanilla autoencoder which only aims to minimize $\mathcal{L}_{Rec}$, the KL divergence term $\mathcal{L}_{KL}$ is an effective regularizer which imposes a multivariate Gaussian prior on the latent space and prevents the VAE from essentially ``memorizing'' the data set and learning a trivial identity mapping through a disconnected latent space \cite{doersch2016tutorial}. Training is performed by passing tuples $(\mathbf{A}_i,\mathbf{T}_i)$ through the VAE in mini-batches of size 32 and updating the network parameters with mini-batch gradient descent using the Adam optimizer \cite{kingma2014adam}. The VAE loss $\mathcal{L}_{VAE}$ is typically observed to plateau within 4,000 epochs. We note that the regularization introduced by the KL divergence term $\mathcal{L}_{KL}$ serves to prevent over-fitting and enables us to train over the full set of molecules to be embedded by the VAE.  

\begin{figure}[ht!]
\centering
  \includegraphics[width=1.0\linewidth]{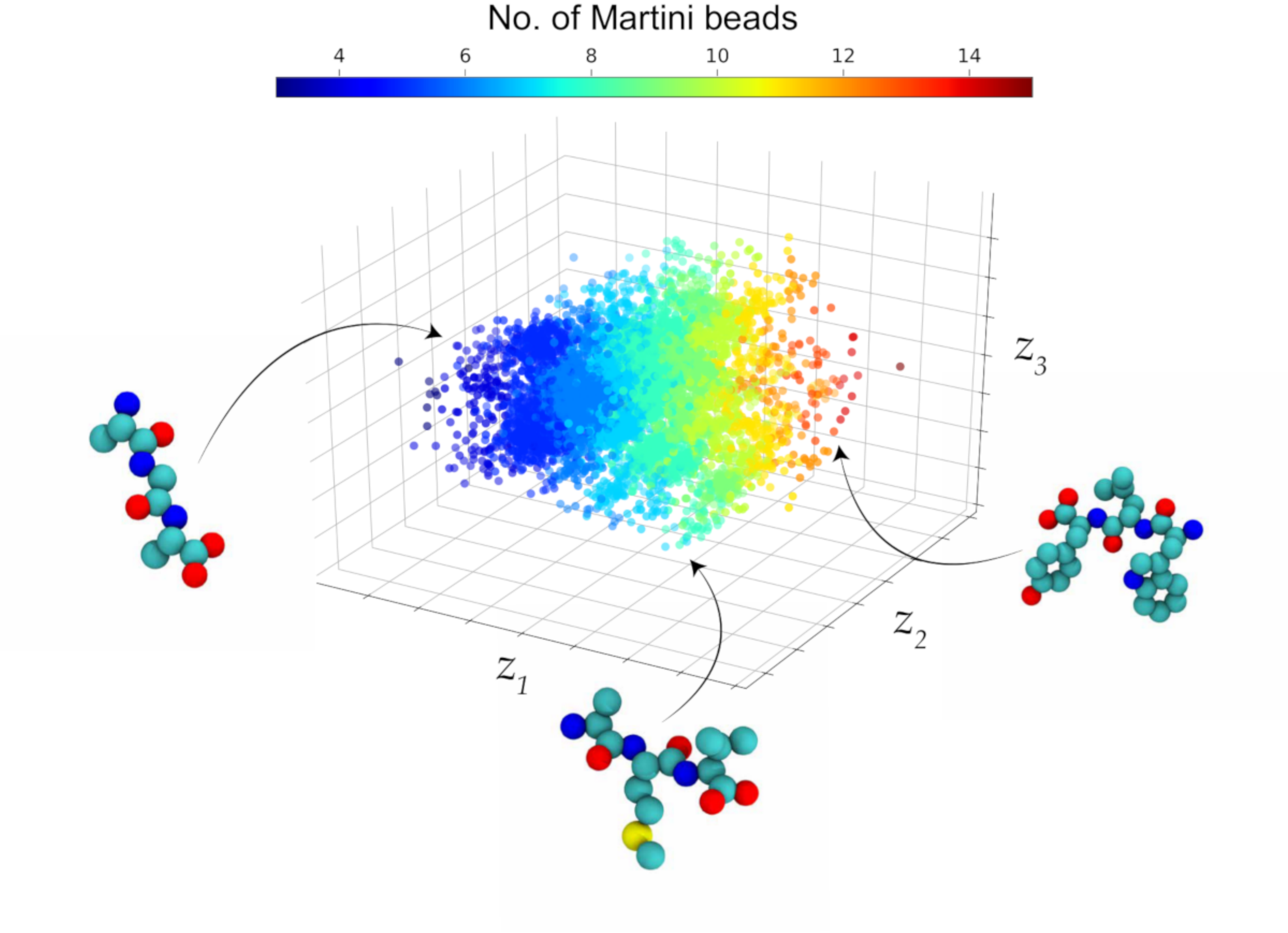}
  \caption{Illustrative visualization of $d = 3$ VAE latent space embedding of the DXXX-OPV3-XXXD family. The embedded molecules are colored according to the number of beads in the XXX tripeptide and selected molecules are visualized. The first dimensional of the latent space $z_1$ is correlated with molecular size $\rho(z_1,\text{size})$ = 0.858 (p-value $<$ 1$\times$$10^{-15}$) providing a visual illustration that similar molecules are embedded close together in the VAE latent space.}
  \label{fgr:3Dembedding}
\end{figure}

We present in Fig.~\ref{fgr:3Dembedding} an example of a $d$ = 3 VAE latent space embedding of the DXXX-OPV3-XXXD family. We color each member of the family by the number of beads in the XXX tripeptide to show that the first dimensional of the latent space $z_1$ is approximately correlated with molecular size, possessing a Pearson correlation coefficient $\rho(z_1,\text{size})$ = 0.858 (p-value < 1$\times$$10^{-15}$). The other two dimensions in this example are also some functions of molecular composition and topology, but prove more challenging to correlate with physically interpretable observables. Physical interpretability of the latent space dimensions is a pleasing but not required property of the embedding. The primary purpose of the VAE embedding is to provide a smooth, low-dimensional molecular representations for the GPR surrogate model. We note that the latent space embedding could be shaped and made more interpretable by simultaneous training of a supervised regression model as suggested by Gomez-Bombarelli et al.~\cite{Gomez-Bombarelli2018AutomaticMolecules}.

\subsubsection{Step 3: Gaussian process regression surrogate models} \label{sec:GPR}

Fitness measurements $f_i, \; i \in sampled$ are available for those molecules DXXX-OPV3-XXXD$_i$ for which we have performed coarse-grained molecular simulation. Given these data we wish to predict the fitness of all remaining candidates $\hat{f}_j, \; j \in unsampled$. This constitutes a supervised regression task where we wish to train a surrogate model $f$ over a small number of training examples to predict the fitness of out-of-training examples as a function of their location in the VAE latent space: $f : \hat{f}_i = f(\mathbf{z}_i) = (f \circ E)(\mathbf{A}_i,\mathbf{T}_i)$. In this manner, the regression model ``short circuits'' expensive direct simulation prediction of fitness with a cheap surrogate model, and eliminates the need to perform exhaustive calculations over all molecules in the family. The quality of the model predictions depends on the number and chemical similarity of the training data: the model is expected to perform better with larger training sets and make more accurate predictions for out-of-training examples that are chemically similar to examples in the training set. As such, we expect the model to improve with additional cycles around the active learning loop. For the purposes of active learning (Section \ref{sec:active}), it is also vital to perform uncertainty quantification on the model predictions so that we can both direct sampling towards the most high-performing candidates predicted by the model (exploitation) and towards undersampled areas where the model possesses the highest uncertainties (exploration) \cite{Brochu2010ALearning}. For this reason, we select Gaussian process regression (GPR) to construct our surrogate model $f : \hat{f}_i = f(\mathbf{z}_i)$ as a flexible, non-parametric, Bayesian regression approach that comes with built-in uncertainty estimates \cite{sivia2006data,ebden2015gaussian,rasmussen2003gaussian,Brochu2010ALearning}.

The fundamental principle of a GPR is to employ a Gaussian process to specify a Bayesian prior distribution over regression functions fitting the data, and then to compute the posterior distribution over those functions that are in agreement with the training data~\cite{rasmussen2003gaussian}. The Gaussian process is fully specified by its mean function, which is typically defined to be zero, and its covariance function for which we choose the popular squared exponential kernel,
\begin{equation}
k(\mathbf{z}, \mathbf{z}^{\prime})=\exp(-\frac{1}{2 \gamma}| | \mathbf{z}-\mathbf{z}^{\prime}| |^{2}),
\end{equation}
where $\mathbf{z}$ and $\mathbf{z}^{\prime}$ denote latent space vectors and the bandwidth of the kernel $\gamma$ is a hyperparameter defining the characteristic length scale over which latent space vectors ``see'' one another. Under these choices, the predicted fitness $f^* = f(\mathbf{z}^*)$ for a new point $\mathbf{z}^*$ outside of the training data is a Gaussian distributed random variable with~\cite{rasmussen2003gaussian,ebden2015gaussian},
\begin{align}
f^* &\sim \mathcal{N}(\mu_{f^*} , \sigma_{f^*}), \\
\mu_{f^*} &= K(\mathbf{z}^*, \mathbf{Z}) \left[ K(\mathbf{Z},\mathbf{Z}) + (\boldsymbol{\sigma}_f^2)^T \mathbf{I} \right]^{-1} \mathbf{f}, \notag \\
\sigma_{f^*} &= K(\mathbf{z}^*,\mathbf{z}^*) - K(\mathbf{z}^*,\mathbf{Z}) \left[ K(\mathbf{Z},\mathbf{Z}) + (\boldsymbol{\sigma}_f^2)^T \mathbf{I} \right]^{-1} K(\mathbf{z}^*,\mathbf{Z})^T, \notag
\end{align}
where $\mathbf{I}$ is the $n$-by-$n$ identity matrix, $\mathbf{f} = \left[ f_1, f_2, \ldots, f_n \right]^T$ is the vector of (noisy) measurements of fitness for the $n$ training points $\mathbf{Z} = \{ \mathbf{z}_1, \mathbf{z}_2, ... , \mathbf{z}_n \}$ computed in our coarse-grained molecular simulations, and $\boldsymbol{\sigma}_f^2 = \left[ \sigma_{f_1}, \sigma_{f_2}, \ldots, \sigma_{f_n} \right]^T$ are associated variances of assumed i.i.d.\ Gaussian noise estimated by block averaging (Section \ref{sec:ObjFunc}), and the $K$ matrices hold the covariances within and between the training data $\mathbf{Z}$ and new point $\mathbf{z}^*$,
\begin{align}
K(\mathbf{Z},\mathbf{Z}) & =
\begin{bmatrix}
k(\mathbf{z}_1, \mathbf{z}_1) & k(\mathbf{z}_1, \mathbf{z}_2) & \cdots & k(\mathbf{z}_1, \mathbf{z}_n) \\
\vdots & \vdots & \ddots & \vdots \\
k(\mathbf{z}_n, \mathbf{z}_1) & k(\mathbf{z}_n, \mathbf{z}_2) & \cdots & k(\mathbf{z}_n, \mathbf{z}_n)
\end{bmatrix}, \\
K(\mathbf{z}^*,\mathbf{Z}) & = \left[ k(\mathbf{z}^*, \mathbf{z}_1), k(\mathbf{z}^*, \mathbf{z}_2), \cdots, k(\mathbf{z}^*, \mathbf{z}_n) \right] , \\
K(\mathbf{z}^*,\mathbf{z}^*) &= k(\mathbf{z}^*,\mathbf{z}^*).
\end{align}
The $(\boldsymbol{\sigma}_f^2)^T \mathbf{I}$ terms account for the uncertainty inherent in our measurements of $\mathbf{f}$ through an assumed Gaussian noise model \cite{ebden2015gaussian}. These terms can also be conceived as a Tikhonov (a.k.a.~ridge or nugget) regularization of the $K(\mathbf{Z},\mathbf{Z})$ matrix that stabilizes its matrix inverse and is particularly useful when this matrix is ill-conditioned due to the close proximity of two or more training points in $\mathbf{Z}$ \cite{mohammadi2016analytic}. A corollary of this regularization is that the GPR posterior is not a perfect interpolator of the training data due to the presence of measurement noise, and we should anticipate residual discrepancies on the order of $\boldsymbol{\sigma}_f$ between the GPR predictions and the our measurements of $\mathbf{f}$. The predictive accuracy and robustness of the GPR is enhanced by the smooth, continuous, and low-dimensional nature of the VAE latent space, which embeds chemically similar points nearby one another and therefore promotes transfer of information to new out-of-training points based on chemically proximate training examples. The GPR prior and posterior are updated during each cycle of the active learning loop as additional training data are collected.

\subsubsection{Step 4: Bayesian optimization} \label{sec:active}

The final step in the cycle is to use the predictions of the surrogate GPR model to identify the next peptide candidates to simulate. We frame this active learning problem as a Bayesian optimization, where we have an expensive, non-differentiable, black-box function with noisy evaluations -- the fitness of each molecule evaluated by coarse-grained molecular simulation -- that we wish to optimize in the minimum number of evaluations. Bayesian optimization defines an acquisition function $u$ that wraps around the current surrogate model to identify peptides with a high chance of being better than the current leader in the training data. We can represent optimization of the acquisition function as,
\begin{align}
\mathbf{z}^\dagger &= \underset{\mathbf{z}}{\mathrm{argmax}} \; u(\mathbf{z} | \mathbf{Z} = \{ (\mathbf{z}_1, f_1), (\mathbf{z}_2, f_2) ... , (\mathbf{z}_n, f_n) \}),
\end{align}
where $\mathbf{z}^\dagger$ is the VAE latent space coordinates of the DXXX-OPV3-XXXD molecule that maximizes the acquisition function $u$, and the maximization is conditioned on the $n$ samples $\{ (\mathbf{z}_1, f_1), (\mathbf{z}_2, f_2) ... , (\mathbf{z}_n, f_n) \}$ collected to date. The surrogate model $f$ enters the maximization through the choice of acquisition function, for which many choices are available~\cite{Brochu2010ALearning}. We employ the popular expected improvement (EI) acquisition function that provides a balanced trade-off between exploitation -- selection of points where the surrogate model posterior mean $\mu_f(\mathbf{z})$ is large -- and exploration -- selection of points where the surrogate model posterior variance $\sigma_f(\mathbf{z})$ is large \cite{Mockus1975OnExtremum,Jones1998EfficientFunctions,Brochu2010ALearning}. Following Lizotte, the EI is defined as~\cite{Lizotte2009PracticalOptimization},
\begin{align}
    u(\mathbf{z} | \mathbf{Z} ) = \operatorname{EI}(\mathbf{z} | \mathbf{Z} ) &=
    \begin{cases} 
      (\mu_f(\mathbf{z})-f(\mathbf{z}^{+})-\xi) \Phi(Z) + \sigma_f(\mathbf{z}) \phi(Z) & \sigma_f(\mathbf{z}) > 0 \\ \label{eqn:EI}
      0 & \sigma_f(\mathbf{z}) = 0 \\
   \end{cases}, \\
    Z &=
    \begin{cases} 
      \frac{\mu_f(\mathbf{z})-f(\mathbf{x}^{+})-\xi}{\sigma_f(\mathbf{z})} & \sigma_f(\mathbf{z}) > 0 \\
      0 & \sigma_f(\mathbf{z}) = 0 \\
   \end{cases},
\end{align}
where $f(\mathbf{z}^{+}), \; \mathbf{z}^{+} \in \mathbf{Z} = \{ \mathbf{z}_1, \mathbf{z}_2, ... , \mathbf{z}_n \}$ is the maximum fitness value among all $n$ sampled candidates to date, $\Phi$ and $\Psi$ are the cumulative distribution function and probability density function of the standard normal distribution, and the hyperparameter $\xi$ controls the exploration-exploitation trade-off. The first term in Eqn.~\ref{eqn:EI} promotes exploitation and the second promotes exploration: when $\xi$ is small the EI will favor exploitation and select points with high posterior mean, while if $\xi$ is large exploration is performed selecting points with large posterior uncertainty~\cite{Brochu2010ALearning}. 

\begin{figure}[ht!]
\centering
  \includegraphics[width=0.9\linewidth]{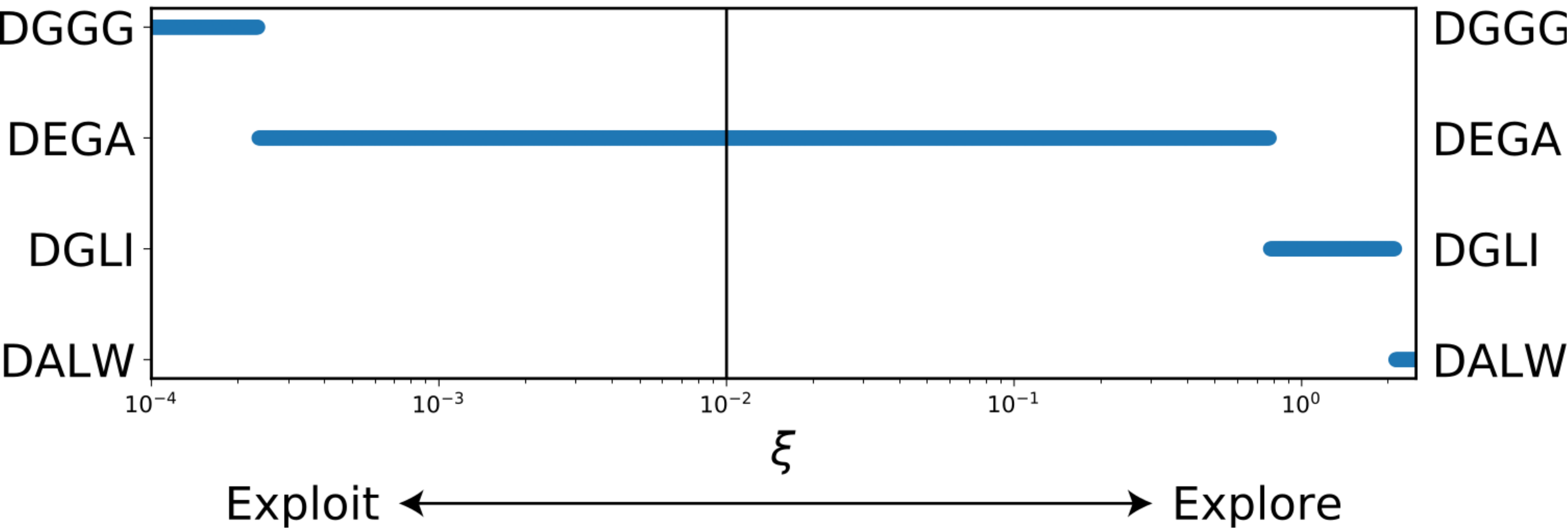}
  \caption{Active learning candidate selection. The expected improvement (EI) acquisition function (Eqn.~\ref{eqn:EI}) is evaluated over for all unsampled members of the DXXX-OPV3-XXXD family at values of the exploit-explore hyperparameter $\xi$ over the range $\log_{10} \xi \in [-4,0.4]$. The blue points in the graph indicate which DXXX candidate maximizes the EI at each value of $\xi$. In this illustrative example there are four candidates -- DGGG, DEGA, DGLI, and DALW -- that maximize the EI over the range of $\xi$ considered. The vertical line shows the recommended value of $\xi = 0.01$ suggested in literature~\cite{Brochu2010ALearning,Lizotte2009PracticalOptimization}, which would result in the selection of only DEGA as the next molecule to simulate. In our approach, we select all four of the molecules DGGG, DEGA, DGLI, and DALW that maximize EI over the entire range of $\xi$ considered as the next best candidates to simulate in parallel in the next round of coarse-grained molecular simulations.}
  \label{fgr:SelectChems}
\end{figure}

Active learning typically proceeds by selecting a fixed value $\xi$ = 0.01~\cite{Brochu2010ALearning,Lizotte2009PracticalOptimization} of the exploration-exploitation trade-off, identifying the candidate that maximizes the EI, and then performing expensive function evaluation (here a coarse-grained molecular simulation) for that candidate. We employ a slightly modified version of this approach that effectively integrates over $\xi$ and performs active learning in batches, which has the advantages of (i) eliminating the sensitivity in selection to the hyperparameter $\xi$, (ii) spreading the exploit-explore trade-off, and (iii) making more efficient use of parallel compute resources to conduct multiple simulations in parallel in the same wall clock time. Specifically, we maximize the EI acquisition function over the range $\log_{10} \xi \in [-4,0.4]$ and select up to four candidates over this range as the ``next-best'' candidates to simulate. Molecules that have already been sampled in preceding rounds are excluded from the pool of available candidates at each round. Where more than four candidates emerge from the EI maximization, we randomly select four members of this set. An example of this selection procedure is presented in Fig.~\ref{fgr:SelectChems}. Coarse grained molecular simulations of these optimal candidates are then performed to commence another round of the active learning cycle.

\subsubsection{Hyperparameter optimization} \label{sec:hyperparameter}

The dimensionality $d$ of the VAE latent space embedding and bandwidth $\gamma$ of the GPR kernel are tunable hyperparameters to be optimized during each cycle of the active learning loop. We perform simultaneous tuning of $d$ and $\gamma$ during each round by creating 50 embeddings of all 8,000 DXXX-OPV3-XXXD molecules into the VAE latent space $\mathbf{z}_i \sim \mathcal{N}(\boldsymbol{\mu}_i , \boldsymbol{\sigma}_i) \in \mathbb{R}^{d}$, each employing different realization of random numbers to sample from the latent space Gaussian for each point, for $d$ = [3, 10]. We then optimize $\gamma$ for each embedding over the range $\gamma$ = [0.001, 100] using a line search followed by Nelder-Mead optimization to maximize the GPR accuracy under cross-validation. We employ leave-one-out cross-validation (LOO-CV) for the first five cycles of the active learning, and then 5-fold CV for subsequent rounds due to the high cost of LOO-CV for larger quantities of samples. The best performing VAE embedding and associated optimal $d$ and $\gamma$ are adopted for the remainder of the current active learning cycle.

\subsubsection{Stop criteria} \label{sec:stop}

We cycle around the active learning loop until the GPR surrogate model no longer improves with the collection of additional training data. A number of stopping criterion for active learning have been proposed \cite{Lorenz2015StoppingOptimization,SchohnLessMachines,Vlachos2008ALearning,Zhu2008Multi-criteria-basedAnnotation,Bloodgood2014AStopping,Bloodgood2015AnalysisPredictions}, but in this work we monitor and define convergence using the stabilizing predictions (SP) method that evaluates performance based on unlabeled data \cite{Bloodgood2014AStopping} and the performance difference (PD) method that considers the labeled examples \cite{Beatty2019TheClassification}. The SP method examines the predictions of consecutive models at each iteration of the active learning procedure on a randomly selected set of 500 points, called the stop set, which is held constant throughout the active learning. We measure the difference in the regression predictions between subsequent rounds using the average Bhattacharyya distance $D_B$\cite{Bhattacharyya1946OnPopulations} between the posterior of consecutive GPR models over the stop set. Large differences in $D_B$ indicate the model is continuing to update the GPR posterior, whereas small values indicate that the surrogate model predictions have stabilized. 

The PD method is used to evaluate model performance by 5-fold CV of the $R^2$ score over the accumulated labeled samples collected to date within each round of active learning. A plateau in the $R^2$ indicates that additional observations result in only marginal improvements to the GPR fit \cite{Beatty2019TheClassification}. A caution in assessing convergence using labeled data is that these data may not be representative of the data as a whole \cite{SchohnLessMachines,Bloodgood2014TakingDatasets}. These concerns are mitigated in our application since our initial data set comprises a set of randomly selected peptides to initialize the active learning procedure and we collect up to four new data points each round across the exploit-explore spectrum to assure broad sampling of chemical space.

\subsection{Nonlinear manifold learning of assembly pathways} \label{sec:DMAPMethods}

We employ diffusion maps as a manifold learning approach to identify the low-dimensional assembly pathways by which the various DXXX-OPV3-XXXD molecules self-assemble into the terminal aggregates. We have previously described the application of diffusion maps to self-assembling systems in Refs.~\cite{Long2018RationalEngineering, Wang2018NonlinearMaterials, Wang2016MesoscaleAggregation, ma2019inverse}. In brief, we compute a distance metric $d(i,j)$ between each pair of interaction graphs $i$ and $j$ harvested from each frame of each molecular simulation trajectory. A number of graph kernels at varying levels of sophistication and abstraction have been proposed to measure the similarity between pairs of graphs \cite{Przulj2007BiologicalDistribution,Aparicio2017TemporalTransitions,Shervashidze2009FastGraphs,Kashima2003MarginalizedGraphs,Bonner2016EfficientFingerprints.}. We follow the approach of Reinhart et al.\ who employed graphlet decompositions as a diffusion map distance metric to analyze colloidal crystallization \cite{Reinhart2018AutomatedMethod}. This approach featurizes a graph by enumerating all topologically unique subgraphs (``graphlets'') with associated node permutations (``orbits'') within the network up to a certain subgraph size (usually up to five vertices), and creating a vector of orbit counts for each vertex in our graph \cite{Hocevar2014ACounting,Przulj2007BiologicalDistribution,Reinhart2018AutomatedMethod}. The vector of orbit counts at each vertex is reweighted to account for over-counting of the smaller graphlets contained in the larger ones (i.e. counts of graphlets comprising two vertices are necessarily contained in counts of graphlets comprising three or more vertices), averaged over all vertices in the graph, and normalized to unit length. This vector represents a featurization of the graph that is permutationally invariant to vertex labeling, and the L2-norm between pairs of vectors defines the graph kernel $d(i,j)$ used to evaluate pairwise distances between our graphical representations of the configurational state of the molecular system.

Diffusion maps then proceed by applying a Gaussian kernel to construct the convolved similarity matrix,
\begin{equation}
A_{i,j}=\exp \left( \frac{- \left( d(i,j)^\alpha \right)^2}{2 \epsilon} \right),
\end{equation}
where the kernel bandwidth $\epsilon$ controls the hop size of the random walk and can be automatically tuned based on the distribution of the $A_{i,j}$ \cite{Coifman2006DiffusionMaps,Wang2016MesoscaleAggregation,Wang2018NonlinearMaterials}. The use of the hyperparameter $\alpha \in (0,1]$ was proposed by Wang et al.\ within a density-adaptive extension of diffusion maps that greatly improves the performance of diffusion maps in applications to systems with large differences in the density of points in the high-dimensional space \cite{Wang2017MesoscaleLandscapes}. For $\alpha$ = 1 we recover standard diffusion maps; for $\alpha \rightarrow 0$ the pairwise distances become increasingly similar and large fluctuations in the density of points in the high-dimensional space are smoothed out. Adopting the tuning procedure proposed in Ref.~\cite{Wang2017MesoscaleLandscapes} we adopt $\alpha$ = 0.15.

The $\bf{A}$ matrix is row normalized to create the right stochastic Markov transition matrix,
\begin{equation}
    \bf{M} = \bf{D}^{-1} \bf{A},
\end{equation}
Where $\bf{D}$ is a diagonal matrix of the row sums of $\bf{A}$.
\begin{equation}
    D_{i,j} = \sum_{j=1}^{N}A_{i,j}.
\end{equation}
The matrix element $M_{i,j}^t=p_t(i,j)$ can interpreted as the probability $p_t(i,j)$ of hopping from point $i$ to point $j$ in $t$ steps of the discrete random walk \cite{Coifman2006DiffusionMaps,Nadler2005DiffusionOperators}. Diagonalization of $\bf{M}$ produces an ordered set of eigenvectors and eigenvalues $\{ (\psi_1 = \vec{1}, \lambda_1 = 1), (\psi_2, \lambda_2), (\psi_3, \lambda_3), ... \}$ with $\lambda_1 = 1 \geq \lambda_2 \geq \lambda_3 \geq \ldots$. The first pair $(\psi_1 = \vec{1}, \lambda_1 = 1)$ is trivial and associated with the stationary distribution of the random walk \cite{Coifman2006DiffusionMaps}. The higher order eigenvectors are associated with a hierarchy of increasingly fast relaxation modes of the random walk. Dimensionality reduction is achieved by identifying a gap in the eigenvalue spectrum after the $\lambda_{k+1}$ to resolve a subspace of slowly relaxing dynamical modes $\{ \psi_2, \psi_3, \ldots, \psi_{k+1} \}$. The diffusion map embedding is the projection of the $i^\text{th}$ interaction graph into the $i^\text{th}$ component of the top $k$ non-trivial eigenvectors,
\begin{equation}
    i \mapsto (\psi_2(i), \psi_3(i),..., \psi_{k+1}(i)).
\end{equation}

We implement this formalism using the memory and compute efficient pivot diffusion map approach that reduces the scaling in the number of points $N$ from $\mathcal{O}(N^2)$ to $\mathcal{O}(N \times n)$, where $n << N$ is the number of so-called ``pivot points'' employed \cite{Wang2018ALearning}. This approach enables the application of diffusion maps to large data sets by performing on-the-fly definition of the $n$ pivot points defining an approximate spanning tree over the high-dimensional data and which are used to support interpolative embeddings of the remaining points.

\section{\label{sec:results}Results and Discussion}

\subsection{Active learning identification of optimal candidates}

The complete DXXX-OPV3-XXXD family comprises 20$^3$ = 8,000 members generated by all permutations of placing each of the 20 natural amino acids within the XXX tripeptide. Prior to conducting active learning we filtered this ensemble to eliminate a subset of candidates containing amino acids known and expected to produce undesired assembly behaviors \cite{Mansbach2018PatchyPeptides}. Specifically, we reduced our search space to the 11$^3$ = 1331 candidates in the set defined by $X \in \{\text{Ala}, \text{Gly}, \text{Glu}, \text{Ile}, \text{Leu}, \text{Met}, \text{Phe}, \text{Trp}, \text{Tyr}, \text{Val}, \text{Asp}\}$ to avoid charged and/or polar amino acids expected to interfere with low-pH triggered assembly \cite{Thurston2018MachineOligopeptides} and focus on those residues that have expressed good assembly behavior in previous experimental work \cite{Wall2014SupramolecularSequence, ardona2015sequence, vadehra2010resin, besar2016organic}.

We perform active learning over DXXX-OPV3-XXXD sequences following the four-part protocol -- molecular simulation, VAE latent space embedding, GPR surrogate model construction, optimal selection of next candidates -- described in Section \ref{sec:activeLearning} and illustrated in Fig.~\ref{fgr:Pipeline}. We seeded the search by conducting coarse-grained molecular dynamics simulations of 90 randomly selected members of the family using the simulation protocol detailed in Section \ref{sec:MD}. This initial broad sampling over the candidate space provides the GPR surrogate model with diverse training data that enables it to identify more- and less-promising regions of the latent space prior to making any predictions. We term this initial round of active learning as Round 0. We conduct 25 additional rounds of active learning (Rounds 1-25) selecting up to four additional molecules for simulation during each pass. This resulted in a sampling a total of $N$ = 186 molecules (2.3\% of the 8,000-member complete family; 14.0\% of the 1331-member chemically restricted family) and a cumulative 558 $\mu$s of simulation time. The particular candidates selected and sampled in each round are listed in \blauw{Table S1} in the \blauw{Supplementary Information}.

\begin{figure}[ht!]
\centering
  \includegraphics[width=\linewidth]{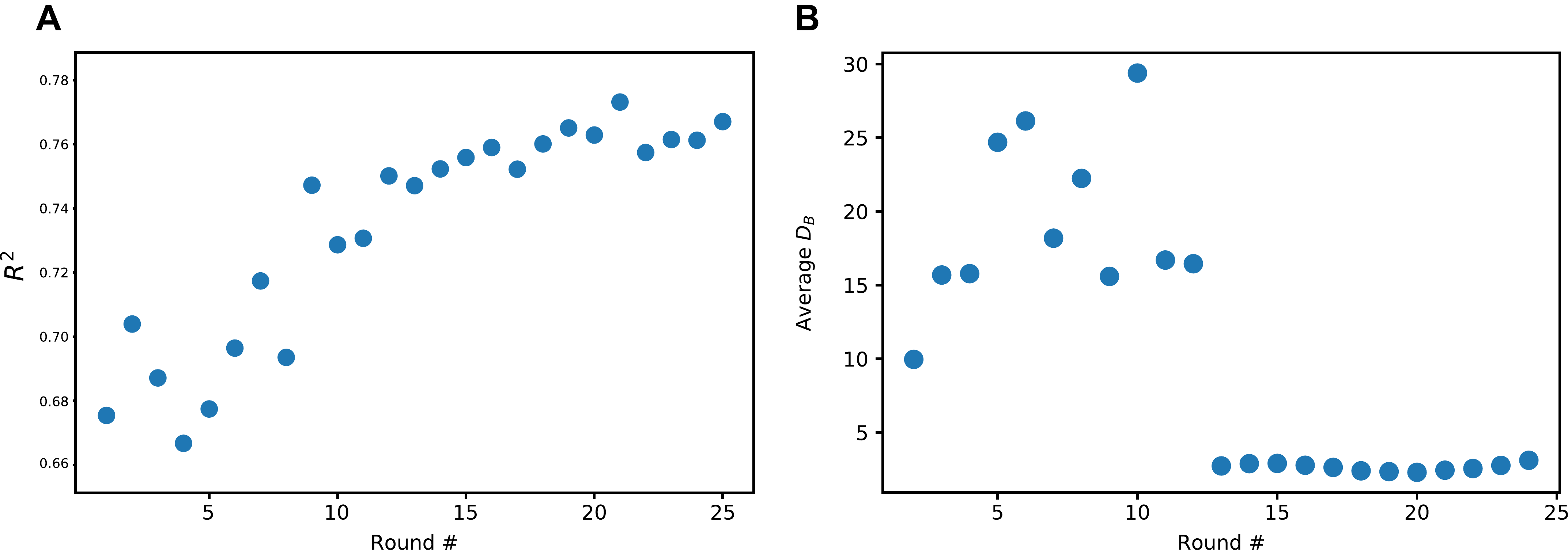}
  \caption{Tracking of active learning stop criteria. (a) The performance difference (PD) method 5-fold cross validation $R^2$ score stabilizes at $R^2$ $\sim$ 78\% by Round 18. (b) The stabilizing predictions (SP) method Bhattacharyya distance between successive GPR posteriors plateaus close to zero at $D_B$ $\sim$ 2.5 after Round 13. Values are reported using a moving average with a window size of three, as recommended in Ref.~\cite{Bloodgood2014AStopping}\ .}
  \label{fgr:Convergence}
\end{figure}

Sampling was terminated by tracking the performance difference (PD) and stabilizing predictions (SP) methods (Section \ref{sec:stop}) \cite{Bloodgood2014AStopping,Beatty2019TheClassification}. The PD method 5-fold cross validation $R^2$ score commences at a reasonably high value of $\sim$68\% -- likely due to the relatively large and diverse $N$ = 90 initial candidates considered, and plateaus to a quite high value of $\sim$78\% by Round 18 (Fig.~\ref{fgr:Convergence}a). The SP method reveals a Bhattacharyya distance between successive GPR posteriors of $D_B$ > 10 over the first 13 rounds, indicating that the additional training data incorporated into the GPR surrogate model are substantially altering its predictions. After Round 14, the Bhattacharyya distance plateaus to $D_B$$\sim$ 2.5 indicating that the surrogate model has stabilized. Rounds 18-25 are therefore proceeding with a stable GPR model and the exploitation candidates identified by the expected improvement acquisition function (Section \ref{sec:active}) furnish the best predictions of the top performing molecules that did not happen to have already been sampled in previous rounds. 

We present in Table \ref{tbl:BestDXXX} the top performing molecules among the 186 that were simulated within our active learning protocol. We report their fitness $f_i = \overline{\kappa}_i$ corresponding to the mean number of $\pi$-core--$\pi$-core contacts per molecule in the terminal self-assembled aggregates (Section \ref{sec:ObjFunc}), the round of active learning in which they were sampled, and whether they have been previously explored by experiment or simulation. Additional molecules that have been previously identified as high performing by experiment and simulation are also presented for comparison. The list of all 1,331 DXXX-OPV3-XXXD molecules in the family with fitness predictions and rankings assigned by the terminal GPR surrogate model is presented in \blauw{Table S2}. There is very good agreement between the numerical simulation results and the GPR model predictions over the training set of 186 molecules for which measurement data exist: the calculated and predicted values of $f_i = \overline{\kappa}_i$ possess a Pearson correlation coefficient of $\rho_\text{Pearson}$ = 0.90 (p-value = 4$\times$$10^{-68}$) and the calculated and predicted rankings possess a Spearman correlation coefficient of $\rho_\text{Spearman}$ = 0.86 (p-value = 1$\times$$10^{-54}$). The agreement is not perfect due to our incorporation of uncertainty estimates in our noisy fitness measurements into GPR training such that the model predictions fall within the error bars of our simulations (Section \ref{sec:GPR}).

\begin{table}[ht!]
\small
  \caption{Top 15 DXXX-OPV3-XXXD molecules identified by the active learning protocol. Additional molecules previously studied in simulation and experiment are also shown for comparison.}
  \label{tbl:BestDXXX}
  \begin{tabular*}{1.0\textwidth}{@{\extracolsep{\fill}}ccccc}
    \hline
    Rank (out of 186) & Molecule (DXXX) & $\overline{\kappa}$ & Discovery round & Previously known? \\
    \hline
    1 & DEAA & 6.06 $\pm$ 0.02 & 1 & N \\
    2 & DDAI & 6.03 $\pm$ 0.02 & 0 & N \\
    3 & DIAM & 6.01 $\pm$ 0.02 & 17 & N \\
    4 & DVAA & 5.95 $\pm$ 0.03 & 9 & N \\
    5 & DAAV & 5.92 $\pm$ 0.03 & 19 & N \\
    6 & DGLG & 5.92 $\pm$ 0.02 & 20 & N \\
    7 & DAEA & 5.92 $\pm$ 0.02 & 25 & N \\
    8 & DAGI & 5.90 $\pm$ 0.01 & 21 & N \\
    9 & DGIG & 5.88 $\pm$ 0.02 & 25 & N \\
    10 & DEAL & 5.88 $\pm$ 0.01 & 23 & N \\
    11 & DGGM & 5.87 $\pm$ 0.04 & 0 & N \\
    12 & DLAV & 5.86 $\pm$ 0.02 & 16 & N \\
    13 & DGDL & 5.85 $\pm$ 0.03 & 0 & N \\
    14 & DGIA & 5.80 $\pm$ 0.04 & 15 & N \\
    15 & DAGL & 5.79 $\pm$ 0.02 & 19 & N \\
    \vdots & \vdots & \vdots & \vdots & \vdots \\
    19 & DVAG & 5.73 $\pm$ 0.01 & 22 & Exp (Ref.~\cite{Wall2014SupramolecularSequence}) \\
    \vdots & \vdots & \vdots & \vdots & \vdots \\
    33 & DAAG & 5.62 $\pm$ 0.01 & 2 & Exp (Ref.~\cite{Wall2014SupramolecularSequence}) \\
    \vdots & \vdots & \vdots & \vdots & \vdots \\
    45 & DGAG & 5.54 $\pm$ 0.03 & 0 & Sim (Ref.~\cite{Mansbach2018PatchyPeptides}); Exp (Ref.~\cite{Wall2014SupramolecularSequence}) \\
    \vdots & \vdots & \vdots & \vdots & \vdots \\
    65 & DFGG & 5.33 $\pm$ 0.03 & 0 &  Exp (Ref.~\cite{Wall2014SupramolecularSequence}) \\
    \vdots & \vdots & \vdots & \vdots & \vdots \\
    85 & DFAV & 5.09 $\pm$ 0.02 & 0 &  Exp (Ref.~\cite{Wall2014SupramolecularSequence}) \\
    \vdots & \vdots & \vdots & \vdots & \vdots \\
    93 & DFAG & 4.98 $\pm$ 0.01 & 0 & Sim (Refs.~\cite{Mansbach2017,Mansbach2017ControlFlow}); Exp (Ref.~\cite{Wall2014SupramolecularSequence}) \\
    \vdots & \vdots & \vdots & \vdots & \vdots \\
    102 & DIAG & 4.86 $\pm$ 0.01 & 2 &  Exp (Ref.~\cite{Wall2014SupramolecularSequence}) \\
    \vdots & \vdots & \vdots & \vdots & \vdots \\
    111 & DFAA & 4.78 $\pm$ 0.01 & 0 &  Exp (Ref.~\cite{Wall2014SupramolecularSequence}) \\
    \vdots & \vdots & \vdots & \vdots & \vdots \\
    147 & DFAF & 4.29 $\pm$ 0.02 & 21 &  Exp (Ref.~\cite{Wall2014SupramolecularSequence}) \\
    \hline
  \end{tabular*}
\end{table}

Trends apparent in the active learning-ranking of the tripeptides in terms of amino acid composition and sequence are coincident with aspects of existing understanding, but also suggest new unexplored amino acid sequences as good putative candidates. The bulky aromatic residues F, W, and Y tend to disfavor good assembly behaviors \cite{Wall2014SupramolecularSequence}, with the large size of these residues impeding good side chain packing and obstructing co-facial stacking of the cores (particularly in the X$_3$ position of DX$_1$X$_2$X$_3$-OPV3-X$_3$X$_2$X$_1$D), and their aromatic character disrupting the formation of linear aggregates with in-register stacking of the $\pi$-cores by introducing favorable aromatic stacking between the $\pi$-cores and peptide wings. These trends are expressed in the low ranking of molecules containing bulky aromatic residues (e.g., DFGG (65), DFAV (85), DFAG (93), DIAG (102), DFAA (111), DFAF(147)) compared to those with smaller hydrophobic side chains (e.g., DVAG (19), DAAG (33), DGAG (45)). The active learning protocol also identifies as highly ranked a number of previously unknown candidates enriched in smaller hydrophobic residues. Interestingly, a number of highly ranked candidates contain an M residue in the X$_3$ position (e.g., DIAM (3), DGGM (11)). Methionene-containing DXXX-OPV3-XXXD molecules have been completely unexplored due, in part, to the expectation that a thioether group would likely disfavor hydrophobic association. Our calculations predict these candidates to possess excellent assembly behaviors and suggest them as novel molecules for experimental investigation.

\subsection{Manifold learning of assembly pathways} \label{sec:DMAP}

The active learning protocol considers only the terminal 50 ns of the 3,000 ns coarse-grained molecular dynamics trajectories to identify DXXX-OPV3-XXXD molecules that form desired pseudo-1D linear aggregates. Having completed the active learning process, we subsequently analyze the ensemble of $N$ = 186 molecular simulation trajectories to provide molecular-level understanding of the assembly pathways and mechanisms and furnish design precepts for the observed assembly behaviors as a function of tripeptide sequence. 

We hypothesize that the molecular assembly trajectories proceed through configurational phase space over a low-dimensional manifold. We determine this low-dimensional manifold by performing diffusion map manifold learning over the trajectory ensemble \cite{Coifman2006DiffusionMaps,Nadler2005DiffusionOperators}. Each frame of each molecular simulation is represented as an interaction graph with vertices $V$ and edges $E$ defined using the optical distance metric (Section \ref{sec:ObjFunc}). We subsample each trajectory keeping every 20$^\text{th}$ point and then run diffusion maps on the composite data set of 558,000 graphs as detailed in Section \ref{sec:DMAPMethods}. Diffusion maps then produce a nonlinear projection of this graph ensemble into a low-dimensional space in which graphs sharing a similar structure of edges are embedded close together, and dissimilar graphs embedded far apart. (We emphasize that this low-dimensional embedding represents a nonlinear manifold residing within the \textit{configurational space} of interaction graphs and is completely independent from the VAE latent space embedding of the \textit{chemical space} of XXX tripeptides.) We trace assembly trajectories over this graph embedding to identify DXXX-OPV3-XXXD molecules that follow similar and dissimilar dynamical assembly pathways and terminal states.

\begin{figure*}[ht!]
 \centering
 \includegraphics[width=\textwidth]{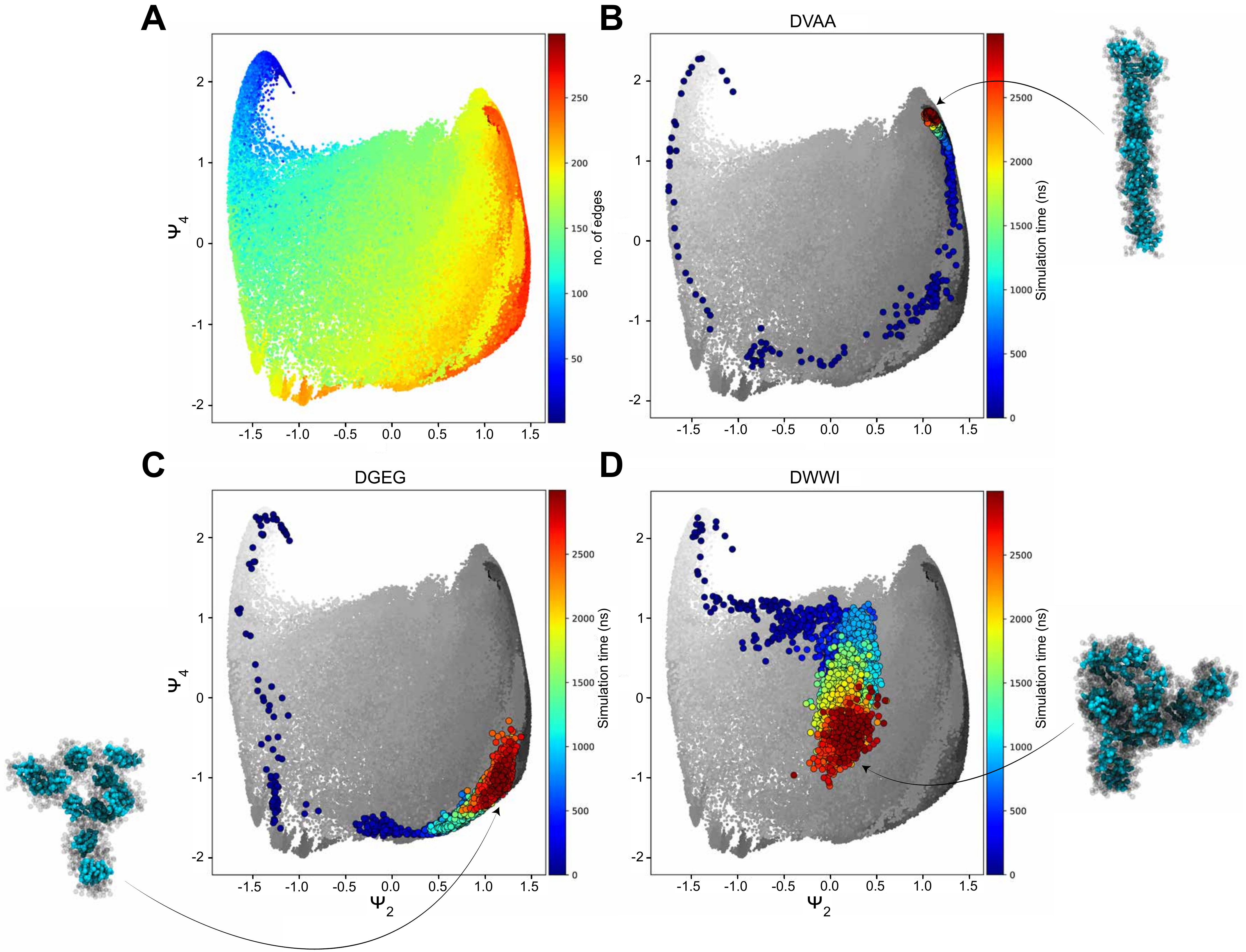}
 \caption{Diffusion map embeddings into $\psi_2-\psi_4$ of the $N$ = 186 DXXX-OPV3-XXXD molecular simulation trajectories. (a) Composite embedding of all 558,000 simulation snapshots. Each point represents a snapshot from one of the simulation trajectories and points are colored by the total number of edges in the corresponding molecular interaction graph (Section \ref{sec:ObjFunc}). Temporal assembly courses of selected molecules over the 2D manifold where points are colored by simulation time: (b) DVAA-OPV3-AAVD (good assembler: $\overline{\kappa}$ = 5.85 $\pm$ 0.03, rank = 4/186), (c) DGEG-OPV3-GEGD (intermediate assembler: $\overline{\kappa}$ = 5.02 $\pm$ 0.02, rank = 89/186), (d) DWWI-OPV3-IWWD (poor assembler: $\overline{\kappa}$ = 3.74 $\pm$ 0.01, rank = 175/186).}
 \label{fgr:DMAP_embed}
\end{figure*}

The diffusion map eigenvalue spectrum possesses a spectral gap after the third non-trivial eigenvalue, motivating 3D embeddings into the three leading eigenvectors $\{ \psi_2, \psi_3, \psi_4 \}$. Further, the $\psi_2-\psi_3$ projection defines a curved, relatively thin manifold indicating that these two embedding dimensions are correlated (\blauw{Fig.~S1}) \cite{Ferguson2010SystematicMaps.}. Accordingly, without too much loss of information we drop $\psi_3$ and construct visually simpler 2D $\psi_2-\psi_4$ embeddings that we present in Fig.~\ref{fgr:DMAP_embed}. In Fig.~\ref{fgr:DMAP_embed}a we present the composite embedding of all 558,000 simulation snapshots. We find $\psi_2$ to be moderately strongly correlated with the average number of per molecule $\pi$-core--$\pi$-core contacts $\kappa$ ($\rho(\psi_2,\kappa)$ = 0.78, p-value < 1$\times$$10^{-15}$), and $\psi_4$ with the mass averaged cluster size of the system $M_z$ ($\rho(\psi_4,M_z)$ = 0.62, p-value < 1$\times$$10^{-15}$) \cite{rubinstein2003polymer}. 

In  Fig.~\ref{fgr:DMAP_embed}b-d we highlight the assembly trajectories for three selected molecules: DVAA-OPV3-AAVD as a good assembler with $\overline{\kappa}$ = (5.85 $\pm$ 0.03) and rank = 4/186, DGEG-OPV3-DGEG as an intermediate assembler with $\overline{\kappa}$ = (5.02 $\pm$ 0.02) and rank = 89/186, and DWWI-OPV3-IWWD as a poor assembler with $\overline{\kappa}$ = (3.74 $\pm$ 0.01) and rank = 175/186 (\blauw{Table S2}). These three examples possess assembly pathways over the manifold that are prototypical of three classes of assembly behavior. All pathways commence in the top-left of the manifold at $(\psi_2 \approx -1.0, \psi_4 \approx 1.8)$ corresponding to an approximate monomeric dispersion. Good assemblers such as DVAA-OPV3-AAVD follow pathways that travel along the lower perimeter of the manifold and terminate in the top-right corner $(\psi_2 \approx -1.5, \psi_4 \approx 1.8) \rightarrow (\psi_2 \approx -1.5, \psi_4 \approx -1.5)  \rightarrow (\psi_2 \approx 1.5, \psi_4 \approx -1.5) \rightarrow (\psi_2 \approx 1.5, \psi_4 \approx 1.5)$. The configurations in the top-right corner comprise pseudo-1D aggregates with good in-register stacking between the $\pi$-cores and large values of $\overline{\kappa}$. Intermediate assemblers such as DGEG-OPV3-DGEG follow similar pathways that traverse the left and bottom perimeter, but terminate in the bottom-right region of the manifold at $(\psi_2 \approx 1.5, \psi_4 \approx -1.5)$. This bottom-right region comprises loosely connected pseudo-1D aggregates which fail to form a globally connected pseudo-1D structure and possess intermediate values of $\overline{\kappa}$. Lastly, poor assemblers such as DWWI-OPV3-IWWD follow pathways that travel along the top of the manifold $(\psi_2 \approx -1.5, \psi_4 \approx 1.0) \rightarrow (\psi_2 \approx 0.5, \psi_4 \approx 1.0)$ and terminate within the bulk of the manifold $(\psi_2 \approx 0.0, \psi_4 \approx -0.5)$ corresponding to disordered aggregates with poor in register stacking and smaller $\overline{\kappa}$.

\subsection{Unsupervised spectral clustering into assembly classes} \label{sec:clust}

\begin{figure}[ht!]
    \centering
    \rotatebox[origin=c]{90}{%
    \begin{minipage}{0.96\textheight}
    \centering
        \includegraphics[width=\linewidth]{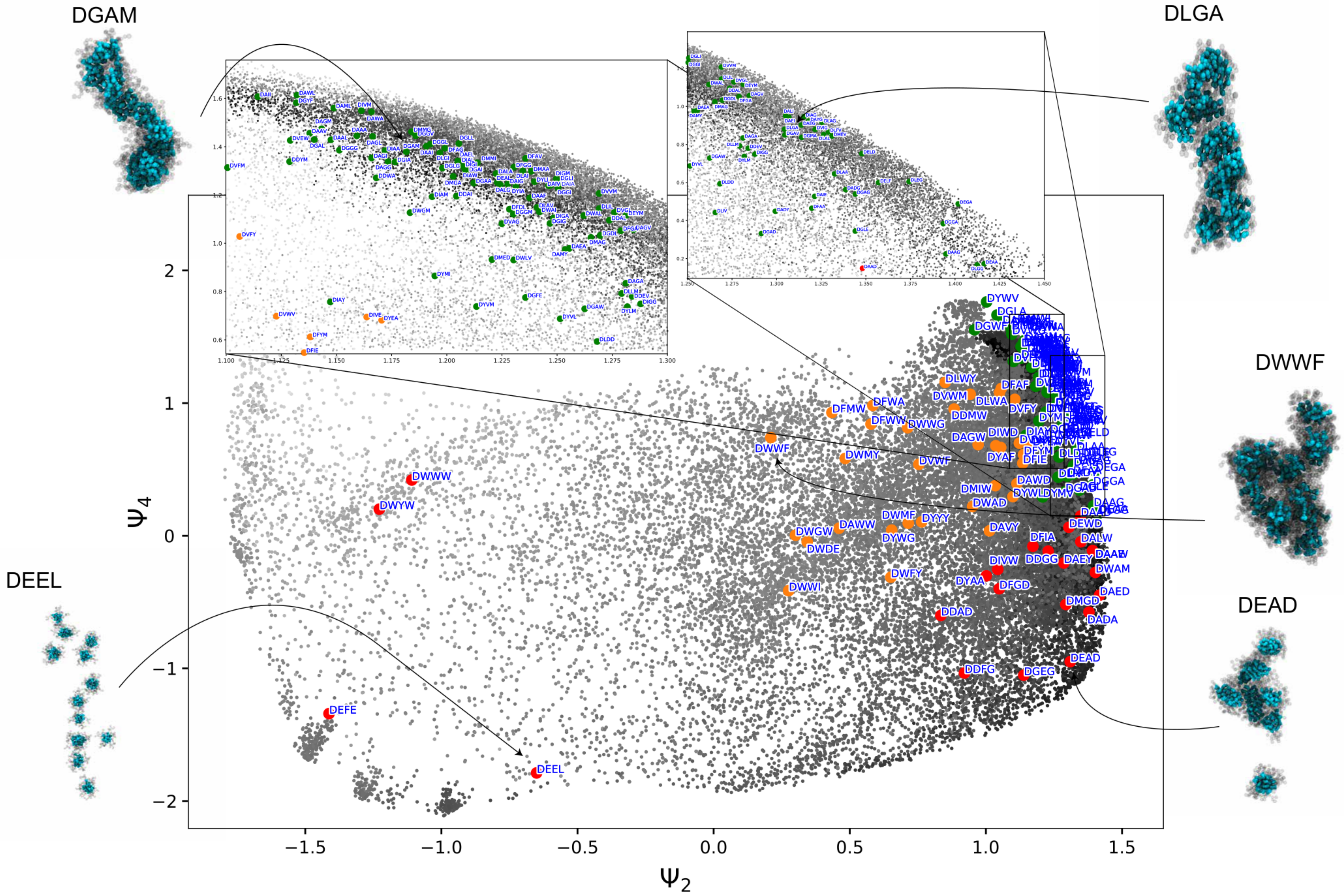}
    \caption{Spectral clustering of the N=186 sampled molecules into four clusters which are projected into the top three non-trivial diffusion map eigenvectors $\{\psi_i\}_{i=2}^{4}$. Points are clustered based on agglomerative hierarchical clustering and cutting the resulting dendrogram at the level of three clusters. Grey points represent instantaneous snapshots harvested from all molecular simulations projected into the $\psi_2$-$\psi_4$ plane. Colored points represent the average coordinates over the terminal 50 ns of simulation for each of the $N$ = 186 molecules. Points are color coded by cluster: green (good assemblers), red (intermediate assemblers), and orange (poor assemblers)}
    \label{fgr:SpectralClusetering}
    \end{minipage}
    }
\end{figure}

The diffusion map embedding of the assembly trajectories presents a means to identify groups of molecules with similar assembly behaviors and extract design precepts to promote good assembly behavior. We map each of the $N$ = 186 DXXX-OPV3-XXXD molecules in the diffusion map embedding to a single 3D point by averaging over the locations of the final 50 ns of simulation data in the space of the top three nontrivial diffusion map eigenvectors $\{\psi_i\}_{i=2}^{4}$. We then perform agglomerative hierarchical clustering using Ward's method \cite{Ward1963HierarchicalFunction}. We cut the resulting dendrogram to partition in the molecules into three clusters (\blauw{Fig.~S2}), and illustrate the clustering of the $N$ = 186 molecules within the $\psi_2-\psi_4$ diffusion map embedding in Fig.~\ref{fgr:SpectralClusetering}. The three clusters reveal a natural categorization into good, intermediate, and poor assemblers: (i) the green cluster of points in the top-right of the embedding comprises the good assemblers that form pseudo-1D linear stacks, (ii) the red cluster located in the bottom-right of the manifold contains intermediate assemblers that form loosely connected small linear aggregates, and (iii) the orange cluster located in the bulk of the manifold that forms disordered and disconnected clusters with poor $\pi$-core stacking. We then propagated the cluster labels defined over these $N$ = 186 molecules to the remaining (1,331 - 186) = 1,145 molecules by performing a nearest-neighbor assignation based on distances within the VAE latent space in the terminal round of active learning (Section \ref{sec:ChemEmbed}). A listing of the cluster assignations of each of the 1,331 molecules is provided in \blauw{Table S3}.

\begin{figure}[ht!]
\centering
  \includegraphics[width=\linewidth]{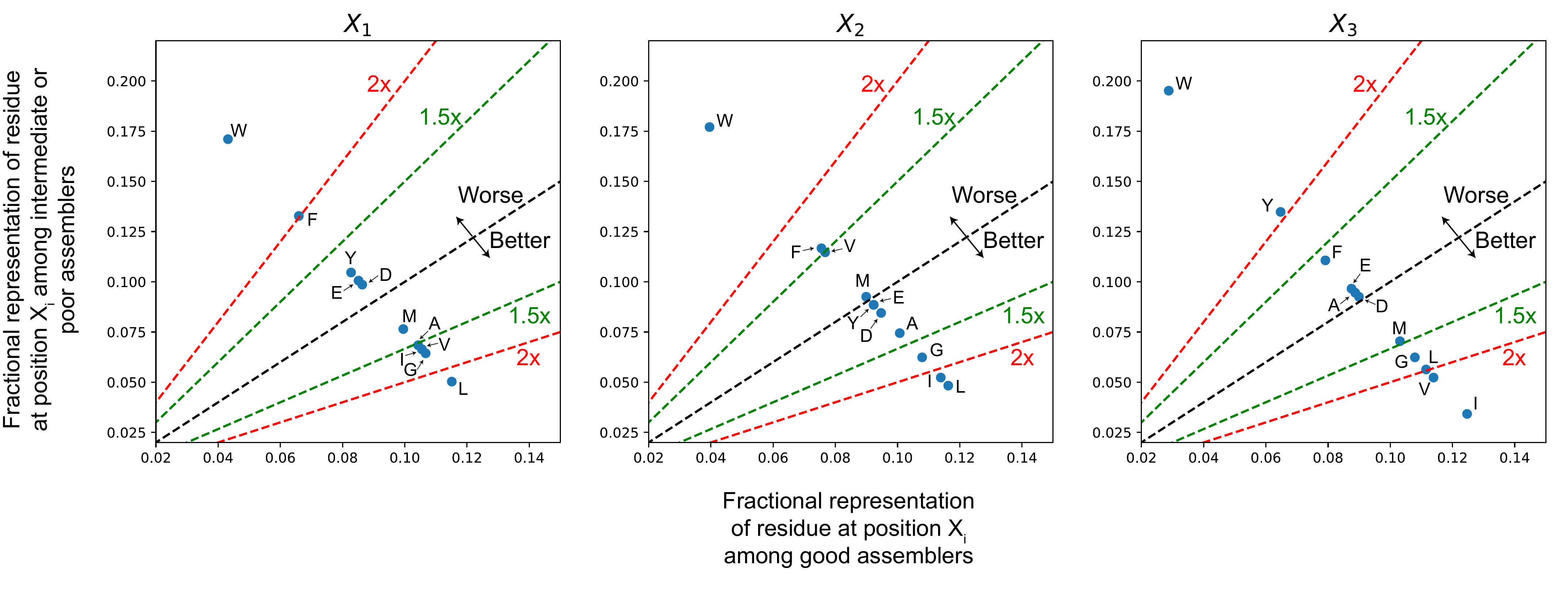}
  \caption{Residue enrichment analysis of each position X$_i$ in DX$_1$X$_2$X$_3$-OPV3-X$_3$X$_2$X$_1$D within molecules classified as good assemblers relative to those classified as intermediate or poor assemblers. Good assemblers are enriched in amino acids residing below the dashed black line and depleted in those residing above it. Dashed green and red lines show boundaries for 1.5$\times$ and 2$\times$ differential enrichment and depletion.}
  \label{fgr:ratios}
\end{figure}

Our classification of the 1,331 molecules allows us to perform a statistical analysis of the enrichment or depletion of amino acid residues in good assemblers relative to intermediate or poor assemblers at each of the the three X$_i$ positions in the DX$_1$X$_2$X$_3$-OPV3-X$_3$X$_2$X$_1$D sequence (Fig.~\ref{fgr:ratios}). A fuller analysis would account for the complete tripeptide sequence to consider the effects of interactions with the other amino acids, but this simpler one-body analysis is both interpretable and illuminating. Drawing a significance cutoff at 1.5$\times$ enrichment or depletion (p-value = 5$\times$$10^{-21}$, one-tailed Fisher's exact test), within good assemblers at the X$_1$ position we observe significant enrichment in $\{ A,G,I,L,V\}$ residues and depletion in $\{F,W\}$. At X$_2$, we observe an enrichment in $\{G,I,L\}$ and impoverishment in $\{F,V,W\}$. Finally, X$_3$ is enriched in $\{G,I,L,V\}$ and impoverished in $\{W,Y\}$. 

First considering the depleted amino acids, the largest hydrophobic residue W is disfavored in good assemblers at all positions. This can be understood as these bulky aromatic side chains possessing favorable $\pi$-stacking interactions with the $\pi$-cores, thereby disrupting $\pi$-core--$\pi$-core stacking. The W residue is most strongly disfavored in the core-adjacent X$_3$ position, where its bulk and proximity to the core can most effectively disrupt good co-facial core stacking. These observations are consistent with the experimental results in Ref.~\cite{Wall2014SupramolecularSequence} where smaller UV-vis spectral shifts were observed upon assembly for molecules containing aromatic residues. Of the remaining two aromatic amino acids, F is similarly disfavored, albeit not to the same degree, but the picture for Y is surprisingly nuanced. Y is moderately disfavored at X$_1$ and strongly disfavored at X$_3$, but at X$_2$ it is neither favored nor disfavored. The latter observation was unanticipated, and we currently lack an understanding for why this should be so. This analysis illuminates how location within the tripeptide acts in concert with the inherent physicochemical attributes of an amino acid to modulate its effect.

In regards to the enriched amino acids, the smaller hydrophobic residues G, I, and L are strongly favored at all positions, with I particularly favored in the X$_3$ position. This preference can be understood as the smaller aliphatic residues enabling closer packing between the peptide wings compared to their bulkier counterparts and their absence of aromatic character reducing interference in the co-facial stacking of $\pi$-cores. Residue A is moderately favored at X$_1$ and X$_2$, but neither favored nor disfavored at X$_3$. Contrariwise, V is moderately to strongly favored at X$_1$ and X$_3$, but moderately disfavored at X$_2$. 

Finally, there is no strong preferences for residues D, E, and M at any of the three positions, with the exception of a moderate favorability for M at position X$_3$. 

\section{Conclusions} \label{sec:concl}

The primary goal of this work was to employ molecular simulation to identify members of the DXXX-OPV3-XXXD oligopeptide family exhibiting promising assembly behaviors into pseudo-1D nanoaggregates with good optoelectronic properties, and to discover design precepts for the good assemblers. Trial-and-error exploration of the full chemical space is computationally and experimentally intractable, motivating our use of techniques from optimal experimental design and deep representational learning to efficiently traverse the space of XXX tripeptide sequences and minimize the number of expensive molecular simulations required to identify the top candidates. Employing a combination of coarse-grained molecular simulation, variational autoencoders, Gaussian process regression, and Bayesian optimization, we define an iterative active learning protocol that constructs surrogate models of assembly behavior based on the simulation data collected to date, and uses these models to optimally direct the next round of simulations. The loop is terminated when the surrogate model ceases to improve with additional simulation data and we can reliably predict the top performing molecules. Using this platform we compute a converged rank ordering of the DXXX-OPV3-XXXD oligopeptides in terms of assembly quality after directly simulating only 2.3\% of all possible oligopeptide sequences. The calculated ranking list is consistent with existing understanding of what constitutes good and bad sequences for assembly, but also reveals new promising candidate molecules as superior assemblers that have not previously been considered. Our ranked list presents an inexpensive filtration of the complete DXXX-OPV3-XXXD sequence space to direct expensive experimental synthesis and characterization efforts towards the most promising candidate molecules.

A subsequent analysis of the molecular simulation trajectories reveals a low-dimensional manifold within the high-dimensional configurational space over which assembly proceeds. Clustering of the simulation trajectories within this space reveals a natural partitioning of the DXXX-OPV3-XXXD family into good, intermediate, and poor assemblers. Statistical analysis of these classes reveals the good assemblers to be enriched in small and intermediate-sized hydrophobic residues, depleted in large aromatic residues, and that Asp, Glu, and Met do not strongly influence the quality of assembly. The one exception to the latter result is that Met in the X position closest to the $\pi$-core does moderately to strongly favor assembly. These design precepts provide understanding of the rankings established by the active learning protocol.  

In sum, this work offers a comprehensive investigation of the assembly landscape of the DXXX-OPV3-XXXD family of $\pi$-conjugated peptides using Bayesian optimal experimental design to guide expensive coarse-grained molecular simulations over microsecond time scales. Our calculations efficiently furnish a rank ordering of the DXXX-OPV3-XXXD and identify a small number of top-performing candidates. While these predictions are only as good as the accuracy of the (coarse-grained) molecular model, they are consistent with existing physicochemical understanding, and can be viewed as a coarse computational filtration of the complete sequence space that can guide subsequent computation and experiment towards the most promising candidates. Ongoing experimental work will attempt to synthesize and test the optoelectronic properties of the candidates in this work, while future computational studies will generalize the approach to D(X)$_n$-$\Pi$-(X)$_n$D molecules by extending the considered chemical space to include different $\Pi$ cores, such as perylenediimide (PDI) or oligothiophene (OT), and varying the length of the peptides. Our platform is also generically extensible to the design of other peptide and peptide-like systems, including antimicrobial peptides, cell-penetrating peptides, intrinsically disordered proteins, and peptoids, where the efficient traversal of chemical space, identification of small numbers of top-performing candidates, and exposure of comprehensible design precepts are prioritized.

\section*{Data Availability}

The coarse-grained molecular simulation trajectories of the self-assembly of the 186 DXXX-OPV3-XXXD molecules conducted in this work are hosted for free public download at the Materials Data Facility~\cite{shmilovich2019mdf}, a project affiliated with the NIST Center for Hierarchical Materials Design \cite{blaiszik2016materials,blaiszik2019data} at \url{http://dx.doi.org/10.18126/xqiz-hzc2}. Python 3 Jupyter notebooks implementing our active search procedures are available on GitHub at \url{https://github.com/KirillShmilovich/ActiveLearningCG}.

\section*{Acknowledgments}
	
This material is based upon work supported by the National Science Foundation under Grant Nos.\ DMR-1841807 and DMR-1728947, and a National Science Foundation Graduate Research Fellowship to K.S. under Grant No.\ DGE-1746045. This work was completed in part with resources provided by the University of Chicago Research Computing Center. We gratefully acknowledge computing time on the University of Chicago high-performance GPU-based cyberinfrastructure (Grant No.\ DMR-1828629). We thank Dr.\ Ben Blaiszik for his assistance in hosting our simulation trajectories on the Materials Data Facility.

\clearpage
\newpage

\bibliography{references}


\clearpage
\newpage

\section*{TOC Graphic}
\begin{figure*}[ht!]
\includegraphics[width=0.45\textwidth]{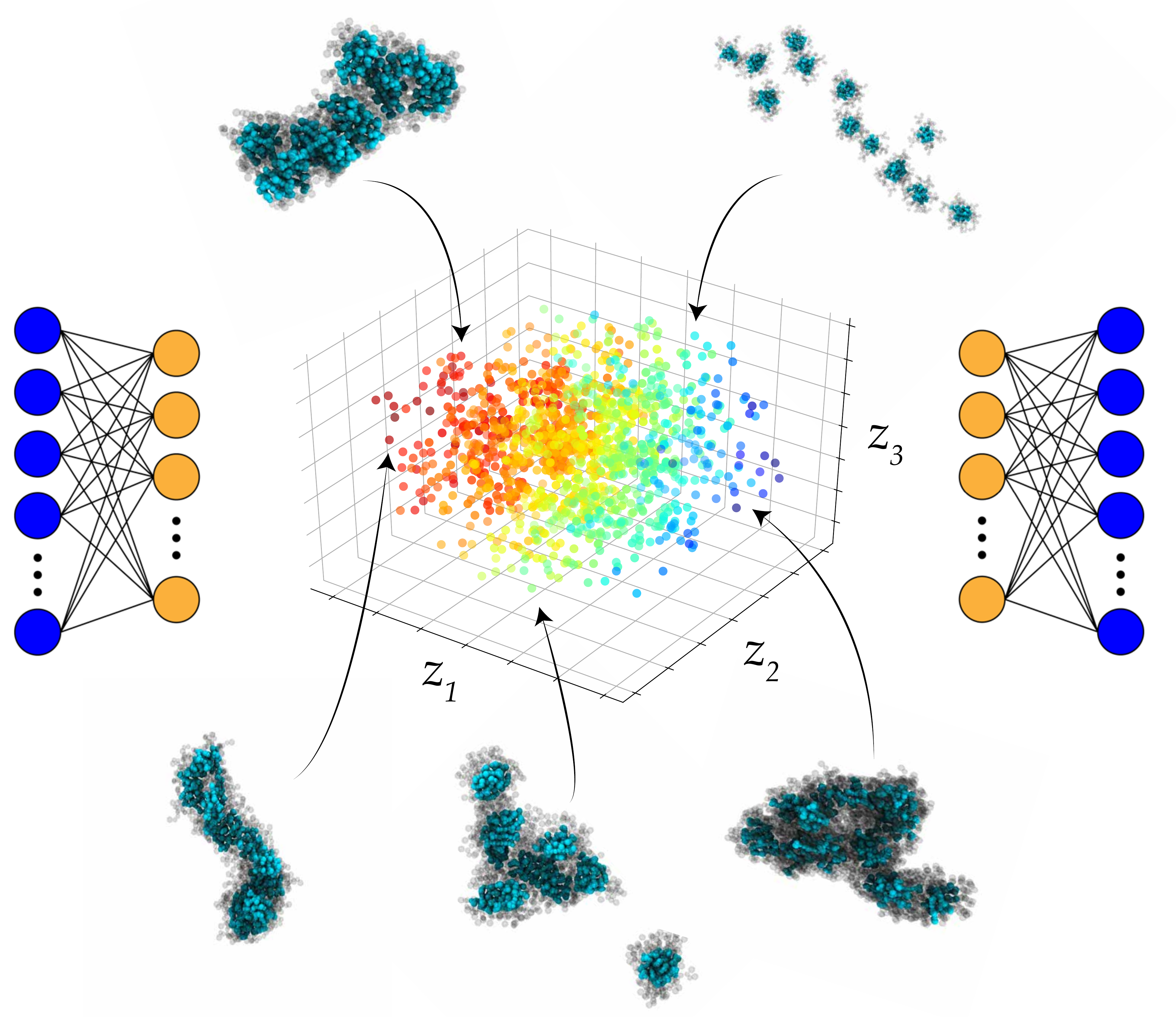}
\caption*{}
\end{figure*}

\end{document}